\documentclass[article,12pt]{elsarticle}

\usepackage{graphicx}

\usepackage{amssymb}
\usepackage{longtable}

\usepackage{amsthm}

\usepackage[nodots,nocompress]{numcompress}

\usepackage{bm}

\journal{JQSRT}

\begin{document}

\begin{frontmatter}

\title{Theoretical investigation of spectroscopic\\ 
properties of $W^{25+}$}%

\author{A. Alkauskas}%
\author{P. Rynkun}
\author{G. Gaigalas}
\author{A. Kynien\.{e}}
\author{R. Kisielius}
\author{S. Ku\v{c}as}
\author{\v{S}. Masys}
\author{G. Merkelis}
\author{V. Jonauskas\corref{cor1}}
\ead{Valdas.Jonauskas@tfai.vu.lt}

\cortext[cor1]{Corresponding author.}

\address{Institute of Theoretical Physics and Astronomy, Vilnius University,
 A. Go\v{s}tauto 12, LT-01108  Vilnius, Lithuania}


\begin{abstract}
Energy levels and emission spectra of $W^{25+}$ ion have been studied by 
performing the large-scale relativistic configuration interaction calculations. 
Configuration interaction strength is used to determine the configurations 
exhibiting the largest influence on the $4f^{3}$,  $4d^{9}4f^{4}$, $4f^{2}5s$, 
$4f^{2}5p$, $4f^{2}5d$, $4f^{2}5f$, $4f^{2}5g$, and $4f^{2}6g$ configuration 
energies. It is shown that correlation effects are crucial for the
$4f^{2}5s \rightarrow 4f^{3}$ transition which in single-configuration approach 
occurs due to the weak electric octupole transitions. As well, the correlation 
effects affect the $4f^{2}5d \rightarrow 4f^{3}$ transitions by increasing 
transition probabilities by an order. Corona model has been used to estimate the 
contribution of various transitions to the emission in a low-density 
electron beam ion trap (EBIT) plasma. Modeling in 10--30 nm wavelength range 
produces lines which do not form emission bands and can be observed in EBIT 
plasma.
\end{abstract}


\end{frontmatter}

\clearpage

\section{Introduction}
Tungsten features many essential properties, such as high-energy threshold of 
sputtering, low sputtering yield, and excellent thermal properties  which make 
it a primary candidate for facing material of walls in thermonuclear reactors. 
However, some tungsten ions as impurities penetrate central regions of 
discharge, and their radiation takes away crucially important energy from fusion 
reaction. Therefore, there is a need of reliable atomic data for all tungsten 
ions in order to monitor the concentration of tungsten in the fusion plasma. 

The aim of the current work is to study the energy levels and emission spectra 
of $W^{25+}$ by performing the large-scale relativistic configuration 
interaction (CI) calculations. The main focus is  put on the emission in the 
spectral range of 2--30 nm, where the intensive radiation of tungsten ions has 
been theoretically and experimentally investigated in fusion, electron-beam ion 
trap (EBIT), and laser-produced plasmas by many authors 
\cite{1977pla_63_295_isler, 1978pla_66_109_hinnov, 1988pla_127_255_Finkenthal, 
1998adndt_68_1_fournier, Biedermann2001PS_T92_85, Radtke2001PhysRevA_64_012720, 
2002cjp_80_1503_utter, Putterich2005_38_3071, 2007jpb_40_3861_ralchenko, 
Jonauskas2007jpb_40_2179, 2007pfr_2_s1060_chowdhuri, 
2008ppcp_50_085016_Putterich, 2010jpb_43_205004_harte, 
2010jpb_43_144009_clementson, 2011pra_83_032517_ralchenko, 
2011jpb_44_175004_suzuki, 2011cjp_89_591_podpaly, 2012adndt_98_557_Bogdanovich, 
2013adndt_99_580_Bogdanovich, 2013epjd_67_1_madeira}. The number of the tungsten 
emission studies is too large to cite here in full. Detailed reviews of
available results for the tungsten spectra and spectral lines in all ionization 
stages are presented in \cite{2009adndt_95_305_kramida, 2011cjp_89_551_kramida}. 

To the best of our knowledge, no fully relativistic with an extended CI basis 
theoretical investigation of emission spectra from $W^{25+}$ ion has been 
carried out so far.The emission from $W^{25+}$ has no detailed theoretical 
studies because of complicated calculations related to the open $f$ shells. 
For ions having configurations with the open $f$ shells, correlation effects may 
play a crucial role due to the mixing of large number of levels. To estimate 
the influence of correlation effects, the configuration interaction strength  
introduced previously in \cite{Karazija1, 1997ps_55_667_kucas} is employed in 
the current work. The same approach has been successfully used for the 
theoretical investigation of transitions in tungsten 
\cite{2010pra_81_012506_jonauskas, 2012adndt_98_19_jonauskas}, transition metals 
\cite{2008litjp_48_219_kyniene}, and even the cascade of Auger processes 
\cite{2008jpb_41_215005_jonauskas, 2010pra_82_043419_palaudoux, 
2011pra_84_053415_jonauskas}. 

The relativistic effects in the Dirac-Fock approach with quantum electrodynamic 
corrections have to be included to provide accurate wavelengths and 
radiative transition probabilities for the highly charged ions of tungsten. 
Previous theoretical investigations included mainly higher than $W^{27+}$  or 
lower than $W^{10+}$ ionization stages of the tungsten ions. Spectra of 
$W^{25+}$ ion have been studied in \cite{2008ppcp_50_085016_Putterich},  
\cite{2010jpb_43_205004_harte}, and  \cite{2011jpb_44_175004_suzuki} where the  
pseudorelativistic approach with scaled integrals 
\cite{Cowan_1981tass.book.....C} is employed. In \cite{2010jpb_43_205004_harte, 
2011jpb_44_175004_suzuki}, the configuration interaction effects among the 
levels of up to three configurations are investigated. There are only a few 
works where energy levels and radiative transition probabilities for the 
tungsten ions with more than 3 electrons in $f$ shell have been studied in the 
Dirac-Fock approach \cite{2010pra_82_014502_gaigalas, 2011pra_83_032509_gaigalas}.

In the following section we describe the multiconfiguration Dirac-Fock method 
used to calculate the energy levels and radiative transition probabilities. 
In Sec. III, the obtained results for the energy levels and emission spectra 
are discussed, and in Sec. IV, the spectra from corona model are presented and 
discussed.

\section{Method of calculation}

The multiconfiguration Dirac-Fock (MCDF) method implemented in the GRASP2K code 
\cite{jonsson2013cpc} adopts the Dirac-Coulomb-Breit Hamiltonian 
\begin{equation}
H^{\mathrm {DCB}} = \sum\limits_{i}\ h_{i}^{\mathrm D} + 
\sum\limits_{i<j}\ h_{ij}^{\mathrm {e}} +
\sum\limits_{i<j}\ h_{ij}^{\mathrm {trans}} ,
\end{equation} 
to solve atomic structure problem for atoms and ions. Here 
$h_{i}^{\mathrm D} = c \alpha_i \cdot {p}_i + (\beta_i - 1) c^{2} + V(r_{i})$ 
is one-electron Dirac Hamiltonian, $ \alpha$ and $\beta$ are the fourth-order 
Dirac matrices, ${p}$ is the momentum operator, $V(r_{i})$ is the electrostatic 
electron-nucleus interaction, $h_{ij}^{\mathrm e}$ is the instantaneous Coulomb 
repulsion and $h_{ij}^{\mathrm {trans}}$ is the transverse interaction operator:
\begin{equation}
h_{ij}^{trans}=\frac{1}{r_{ij}}\left[ 1-\alpha _{i}\cdot \alpha _{j}~\cos (\omega
r_{ij})+(\alpha _{i}\cdot \nabla _{i})(\alpha _{j}\cdot \nabla _{j})~\frac{%
\cos (\omega r_{ij})-1}{\omega ^{2}}\right] 
\end{equation}
where $\omega$ is the energy of a single photon exchanged between a pair of 
electrons $i$ and $j$. 

Diagonalization of the Dirac-Coulomb-Breit Hamiltonian produces atomic state 
functions, $\Psi_{\gamma}$, which characterize different fine-structure states 
and are expressed through the configuration state functions (CSFs)
\begin{equation}
\Psi_{\gamma} (J)=\sum\limits_{\zeta}c_{\gamma}(\zeta J)~\Phi (\zeta J) , 
\label{McdfPsi} 
\end{equation}
where $\zeta$ represents any additional quantum numbers required to uniquely 
specify a state. The CSFs, $\Phi (\zeta J)$,  are built from a basis of 
two-component orbitals
\begin{equation}
\phi (r)=\frac{1}{r}\left( 
\begin{array}{rc}
P_{nlj}(r) & \chi _{ljm}(\hat{r}) \\ 
{\mathrm i}~Q_{n\bar{l}j}(r) & \chi _{\bar{l}jm}(\hat{r})%
\end{array}%
\right). 
\end{equation} 
Here $P_{nlj}(r)$ and $Q_{n\bar{l}j}(r)$ are the large and small radial 
components of one-electron wavefunctions and $\chi _{ljm}(\hat{r})$ are the 
two-component Pauli spherical spinors.

QED corrections, which include vacuum polarization and self-energy (known as 
the Lamb shift), are considered in the first order of perturbation theory. 
Finally, the effects of finite nuclear size are modeled by using two-component 
Fermi statistical distribution function.

\section{Energy levels and radiative transition probabilities}
\label{sec:gA}


\begin{figure}
 \includegraphics[scale=0.5]{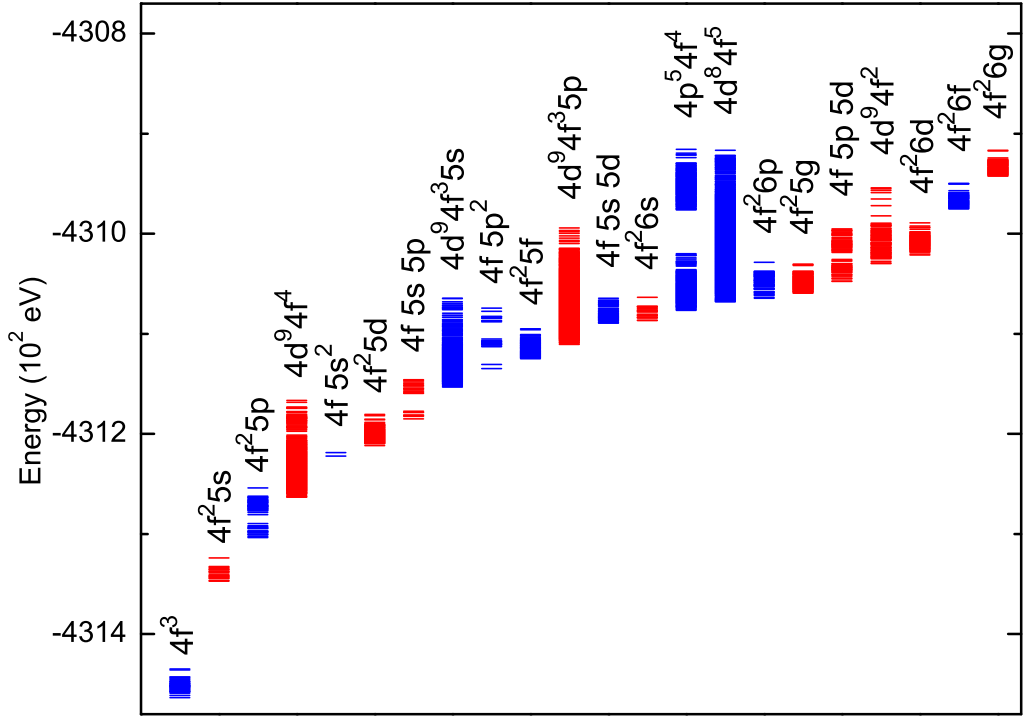}%
 \caption{\label{w25energy} Energy levels of studied configurations in 
$W^{25+}$. }
\end{figure}

Theoretical energy levels for 22 configurations of $W^{25+}$ are presented in 
Figure \ref{w25energy}. The levels are identified by their dominant components 
in the $LS$ coupled basis set. We include all the configurations which have 
energy levels either lower or overlaping with the highest energy level of 
$4d^{9}4f^{3}5s$ configuration. In addition, the configurations which correspond 
to the one-electron excitations from the ground $4f^{3}$ configuration up to 
$6g$ shell are also presented. The configuration $4d^{9}4f^{3}5s$ originates 
from the ground configuration $4f^{3}$ of the ion after the one-electron 
excitation from the $4d$ to $5s$ shell. The total number of levels  in the
present work equals to 13937. The ground configuration of the ion consists of 
41 levels (Table \ref{t1}). It can be seen from Table \ref{t1} that a mixing 
among the configuration state functions in $LS$-coupling is very prominent for 
some levels of the ground configuration. On the other hand, our expansion 
coefficients for the atomic wavefunctions in $jj$-coupling (not presented here) 
show that none of the coupling schemes are suitable for the unique 
identification of some levels. 

The first excited configuration $4f^{2}5s$ can decay to the ground configuration 
only through the electric octupole transitions in single-configuration 
approach. The largest probabilities for these transitions are of $10$ s$^{-1}$ 
order, and the transitions concentrate in the 10--12 nm region. Configuration 
mixing opens decay channels for electric dipole transitions to the levels of 
the ground configuration. However, their probabilities are much weaker compared 
to the other considered electric dipole transitions. The largest transition 
probabilities are of $10^{7}$ s$^{-1}$ order. The analysis of percentage 
compositions of the configuration state functions for levels reveals that the 
mixing of $4f^{2}5s$  with $4d^{9}4f^{4}$, $4f^{2}5d$, and $4d^{9}4f^{3}5p$ 
configurations is responsible for the  electric dipole transitions. However, 
the percentage composition of the configuration state functions from these three 
configurations does not reach more than 1\%. It should be noted that a very 
large CI basis is needed for such a small mixing of the configuration state 
functions in order to provide the reliable values of the radiative transition 
probabilities. The lines from $4f^{2}5s \rightarrow 4f^{3} $ transition appear 
in the spectra when the population of the higher levels goes down through the 
steps of radiative transitions. Likewise, such process can appear on the scene 
in EBIT plasma when the interaction with electrons ends for ions ejected from 
electron beam \cite{2013jqsrt_jonauskas} or in magnetic trapping mode 
\cite{1996rsi_67_11_Beiersdorfer}. The charge exchange of the ions with neutrals 
can also result in the radiative cascade of transferred electrons from the 
highly excited states of the ions \cite{2000prl_85_5090_beiersdorfer, 
2005pra_72_032725_beiersdorfer}.


\begingroup 

\renewcommand{\arraystretch}{1.2}
\renewcommand{\tabcolsep}{1mm}
\scriptsize

\begin{flushleft}

\LTcapwidth 12cm
\begin{longtable}{  @{}r r r l }
\caption{\label{t1} 
MCDF energy levels  with spectroscopic identifications and leading percentage 
composition of the wave functions in  $W^{25+}$. Levels having the largest 
lifetimes and levels to which radiative transition takes place from these 
levels are presented. $J$P stands for the total angular momentum quantum
number $J$ and parity P. Energies are given relative to the ground energy 
$E_{ground}$ = $-$15855.9891992 a.u. 
 }
\\ 
\hline
 Index &     $J$P & $E$ (a.u.) &     Composition \\ 
\hline
\endfirsthead
\caption[]{ (continued) } \\
\hline
 Index &     $J$P & $E$ (a.u.) &     Composition \\ 
\hline
\endhead
\hline 
\multicolumn{4}{r}{\textit{Continued on next page}} \\
\endfoot
\hline
\endlastfoot
     1 &  9/2$-$   & $E_{ground}$    &    82\%  4f$^3$ ($_1^4$I) $^4$I
                                      $+$ 14\%  4f$^3$ ($_2^2$H) $^2$H
                                      $+$  2\%  4f$^3$ ($_1^2$H) $^2$H
                                      $+$  1\%  4f$^3$ ($_1^2$G) $^2$G
                                      $+$  1\%  4f$^3$ ($_2^2$G) $^2$G  \\
     2 & 11/2$-$   &      0.0902578  &    94\%  4f$^3$ ($_1^4$I) $^4$I $+$  4\%  4f$^3$ ($_2^2$H) $^2$H $+$  1\%  4f$^3$ ($_1^2$H) $^2$H  \\
     3 &  3/2$-$   &      0.1632922  &    64\%  4f$^3$ ($_1^4$F) $^4$F
                                      $+$ 22\%  4f$^3$ ($_1^2$D) $^2$D
                                      $+$  7\%  4f$^3$ ($_1^2$P) $^2$P
                                      $+$  3\%  4f$^3$ ($_2^2$D) $^2$D
                                      $+$  3\%  4f$^3$ ($_1^4$S) $^4$S  \\
     4 & 13/2$-$   &      0.1677164  &    92\%  4f$^3$ ($_1^4$I) $^4$I $+$  6\%  4f$^3$ ($_1^2$K) $^2$K $+$  1\%  4f$^3$ ($_1^2$I) $^2$I  \\
     5 &  9/2$-$   &      0.2012862  &    33\%  4f$^3$ ($_2^2$H) $^2$H $+$ 19\%  4f$^3$ ($_1^2$G) $^2$G $+$ 14\%  4f$^3$ ($_2^2$G) $^2$G $+$ 13\%  4f$^3$ ($_1^4$I) $^4$I $+$ 11\%  4f$^3$ ($_1^4$F) $^4$F  \\
     6 &  5/2$-$   &      0.2240156  &    77\%  4f$^3$ ($_1^4$F) $^4$F $+$ 10\%  4f$^3$ ($_1^2$D) $^2$D $+$  8\%  4f$^3$ ($_1^4$G) $^4$G $+$  2\%  4f$^3$ ($_1^2$F) $^2$F $+$  2\%  4f$^3$ ($_2^2$F) $^2$F  \\
     7 & 15/2$-$   &      0.2350187  &    82\%  4f$^3$ ($_1^4$I) $^4$I $+$ 16\%  4f$^3$ ($_1^2$K) $^2$K $+$  1\%  4f$^3$ ($_1^2$L) $^2$L  \\
     8 &  3/2$-$   &      0.2488323  &    63\%  4f$^3$ ($_1^4$S) $^4$S $+$ 22\%  4f$^3$ ($_1^2$P) $^2$P $+$ 13\%  4f$^3$ ($_1^4$F) $^4$F $+$  1\%  4f$^3$ ($_1^4$D) $^4$D $+$  1\%  4f$^3$ ($_1^2$D) $^2$D  \\
     9 &  7/2$-$   &      0.2506589  &    50\%  4f$^3$ ($_1^4$F) $^4$F $+$ 25\%  4f$^3$ ($_1^2$G) $^2$G $+$ 17\%  4f$^3$ ($_2^2$G) $^2$G $+$  6\%  4f$^3$ ($_1^4$G) $^4$G $+$  1\%  4f$^3$ ($_2^2$F) $^2$F  \\
    10 &  5/2$-$   &      0.2631267  &    80\%  4f$^3$ ($_1^4$G) $^4$G $+$ 10\%  4f$^3$ ($_1^4$F) $^4$F $+$  4\%  4f$^3$ ($_1^2$F) $^2$F $+$  3\%  4f$^3$ ($_2^2$F) $^2$F $+$  1\%  4f$^3$ ($_2^2$D) $^2$D  \\
    11 &  7/2$-$   &      0.3041443  &    63\%  4f$^3$ ($_1^4$G) $^4$G $+$ 25\%  4f$^3$ ($_1^4$F) $^4$F $+$  4\%  4f$^3$ ($_2^2$G) $^2$G $+$  4\%  4f$^3$ ($_1^2$G) $^2$G $+$  2\%  4f$^3$ ($_2^2$F) $^2$F  \\
    12 &  9/2$-$   &      0.3189015  &    61\%  4f$^3$ ($_1^4$F) $^4$F $+$ 26\%  4f$^3$ ($_2^2$H) $^2$H $+$  6\%  4f$^3$ ($_1^2$G) $^2$G $+$  3\%  4f$^3$ ($_1^4$I) $^4$I $+$  1\%  4f$^3$ ($_2^2$G) $^2$G  \\
    13 & 11/2$-$   &      0.3290696  &    66\%  4f$^3$ ($_2^2$H) $^2$H $+$ 19\%  4f$^3$ ($_1^4$G) $^4$G $+$  5\%  4f$^3$ ($_1^2$H) $^2$H $+$  4\%  4f$^3$ ($_1^4$I) $^4$I $+$  4\%  4f$^3$ ($_1^2$I) $^2$I  \\
    14 & 13/2$-$   &      0.3360449  &    88\%  4f$^3$ ($_1^2$K) $^2$K $+$  7\%  4f$^3$ ($_1^4$I) $^4$I $+$  4\%  4f$^3$ ($_1^2$I) $^2$I  \\
    15 &  3/2$-$   &      0.3656872  &    24\%  4f$^3$ ($_1^4$S) $^4$S $+$ 21\%  4f$^3$ ($_1^2$D) $^2$D $+$ 16\%  4f$^3$ ($_1^4$F) $^4$F $+$ 14\%  4f$^3$ ($_1^4$D) $^4$D $+$ 14\%  4f$^3$ ($_1^2$P) $^2$P  \\
    16 &  7/2$-$   &      0.3677222  &    29\%  4f$^3$ ($_1^2$G) $^2$G $+$ 27\%  4f$^3$ ($_1^4$G) $^4$G $+$ 23\%  4f$^3$ ($_1^4$F) $^4$F $+$ 19\%  4f$^3$ ($_2^2$G) $^2$G $+$  1\%  4f$^3$ ($_2^2$F) $^2$F  \\
    17 &  1/2$-$   &      0.3811356  &    65\%  4f$^3$ ($_1^2$P) $^2$P $+$ 33\%  4f$^3$ ($_1^4$D) $^4$D  \\
    18 &  9/2$-$   &      0.3813196  &    84\%  4f$^3$ ($_1^4$G) $^4$G $+$  7\%  4f$^3$ ($_1^2$H) $^2$H $+$  5\%  4f$^3$ ($_2^2$H) $^2$H $+$  1\%  4f$^3$ ($_2^2$G) $^2$G $+$  1\%  4f$^3$ ($_1^4$I) $^4$I  \\
    19 & 15/2$-$   &      0.3974398  &    49\%  4f$^3$ ($_1^2$K) $^2$K $+$ 36\%  4f$^3$ ($_1^2$L) $^2$L $+$ 14\%  4f$^3$ ($_1^4$I) $^4$I  \\
    20 &  5/2$-$   &      0.4100329  &    30\%  4f$^3$ ($_1^4$D) $^4$D $+$ 29\%  4f$^3$ ($_2^2$D) $^2$D $+$ 26\%  4f$^3$ ($_1^2$D) $^2$D $+$  4\%  4f$^3$ ($_1^4$F) $^4$F $+$  4\%  4f$^3$ ($_1^4$G) $^4$G  \\
    21 & 11/2$-$   &      0.4298672  &    46\%  4f$^3$ ($_1^2$H) $^2$H $+$ 35\%  4f$^3$ ($_1^4$G) $^4$G $+$ 13\%  4f$^3$ ($_1^2$I) $^2$I $+$  4\%  4f$^3$ ($_2^2$H) $^2$H $+$  1\%  4f$^3$ ($_1^4$I) $^4$I  \\
    22 &  3/2$-$   &      0.4419216  &    46\%  4f$^3$ ($_1^4$D) $^4$D $+$ 26\%  4f$^3$ ($_2^2$D) $^2$D $+$ 12\%  4f$^3$ ($_1^2$D) $^2$D $+$ 10\%  4f$^3$ ($_1^2$P) $^2$P $+$  4\%  4f$^3$ ($_1^4$S) $^4$S  \\
    23 &  5/2$-$   &      0.4485547  &    59\%  4f$^3$ ($_1^2$D) $^2$D $+$ 26\%  4f$^3$ ($_1^4$D) $^4$D $+$  6\%  4f$^3$ ($_1^4$F) $^4$F $+$  4\%  4f$^3$ ($_2^2$D) $^2$D $+$  1\%  4f$^3$ ($_2^2$F) $^2$F  \\
    24 &  9/2$-$   &      0.4659291  &    33\%  4f$^3$ ($_1^2$G) $^2$G $+$ 25\%  4f$^3$ ($_1^4$F) $^4$F $+$ 17\%  4f$^3$ ($_2^2$H) $^2$H $+$ 17\%  4f$^3$ ($_2^2$G) $^2$G $+$  5\%  4f$^3$ ($_1^2$H) $^2$H  \\
    25 & 11/2$-$   &      0.4768551  &    58\%  4f$^3$ ($_1^2$I) $^2$I $+$ 35\%  4f$^3$ ($_1^4$G) $^4$G $+$  3\%  4f$^3$ ($_2^2$H) $^2$H $+$  3\%  4f$^3$ ($_1^2$H) $^2$H  \\
    26 &  1/2$-$   &      0.4989921  &    65\%  4f$^3$ ($_1^4$D) $^4$D $+$ 33\%  4f$^3$ ($_1^2$P) $^2$P  \\
    27 &  3/2$-$   &      0.5203918  &    30\%  4f$^3$ ($_1^2$P) $^2$P $+$ 24\%  4f$^3$ ($_1^2$D) $^2$D $+$ 21\%  4f$^3$ ($_2^2$D) $^2$D $+$ 15\%  4f$^3$ ($_1^4$D) $^4$D $+$  5\%  4f$^3$ ($_1^4$S) $^4$S  \\
    28 & 13/2$-$   &      0.5316025  &    94\%  4f$^3$ ($_1^2$I) $^2$I $+$  5\%  4f$^3$ ($_1^2$K) $^2$K  \\
    29 &  7/2$-$   &      0.5316294  &    91\%  4f$^3$ ($_1^4$D) $^4$D $+$  5\%  4f$^3$ ($_1^2$F) $^2$F $+$  1\%  4f$^3$ ($_2^2$F) $^2$F $+$  1\%  4f$^3$ ($_1^4$G) $^4$G $+$  1\%  4f$^3$ ($_1^2$G) $^2$G  \\
    30 & 15/2$-$   &      0.5341773  &    62\%  4f$^3$ ($_1^2$L) $^2$L $+$ 34\%  4f$^3$ ($_1^2$K) $^2$K $+$  2\%  4f$^3$ ($_1^4$I) $^4$I  \\
    31 &  9/2$-$   &      0.5486678  &    74\%  4f$^3$ ($_1^2$H) $^2$H $+$ 13\%  4f$^3$ ($_2^2$G) $^2$G $+$  6\%  4f$^3$ ($_1^4$G) $^4$G $+$  4\%  4f$^3$ ($_2^2$H) $^2$H $+$  1\%  4f$^3$ ($_1^4$F) $^4$F  \\
    32 & 17/2$-$   &      0.5545060  &    99\%  4f$^3$ ($_1^2$L) $^2$L  \\
    33 &  5/2$-$   &      0.5863249  &    33\%  4f$^3$ ($_1^4$D) $^4$D $+$ 30\%  4f$^3$ ($_2^2$F) $^2$F $+$ 20\%  4f$^3$ ($_1^2$F) $^2$F $+$  9\%  4f$^3$ ($_2^2$D) $^2$D $+$  5\%  4f$^3$ ($_1^4$G) $^4$G  \\
    34 &  3/2$-$   &      0.6041386  &    39\%  4f$^3$ ($_2^2$D) $^2$D $+$ 23\%  4f$^3$ ($_1^4$D) $^4$D $+$ 19\%  4f$^3$ ($_1^2$D) $^2$D $+$ 16\%  4f$^3$ ($_1^2$P) $^2$P $+$  1\%  4f$^3$ ($_1^4$S) $^4$S  \\
    35 & 11/2$-$   &      0.6294527  &    44\%  4f$^3$ ($_1^2$H) $^2$H $+$ 24\%  4f$^3$ ($_1^2$I) $^2$I $+$ 21\%  4f$^3$ ($_2^2$H) $^2$H $+$  9\%  4f$^3$ ($_1^4$G) $^4$G  \\
    36 &  7/2$-$   &      0.6883423  &    65\%  4f$^3$ ($_2^2$F) $^2$F $+$ 21\%  4f$^3$ ($_1^2$F) $^2$F $+$  6\%  4f$^3$ ($_1^4$D) $^4$D $+$  4\%  4f$^3$ ($_1^2$G) $^2$G $+$  2\%  4f$^3$ ($_1^4$G) $^4$G  \\
    37 &  5/2$-$   &      0.6932423  &    56\%  4f$^3$ ($_2^2$D) $^2$D $+$ 17\%  4f$^3$ ($_1^2$F) $^2$F $+$ 13\%  4f$^3$ ($_2^2$F) $^2$F $+$ 10\%  4f$^3$ ($_1^4$D) $^4$D $+$  2\%  4f$^3$ ($_1^4$G) $^4$G  \\
    38 &  9/2$-$   &      0.7512285  &    52\%  4f$^3$ ($_2^2$G) $^2$G $+$ 40\%  4f$^3$ ($_1^2$G) $^2$G $+$  5\%  4f$^3$ ($_1^2$H) $^2$H $+$  1\%  4f$^3$ ($_2^2$H) $^2$H  \\
    39 &  7/2$-$   &      0.7696339  &    53\%  4f$^3$ ($_2^2$G) $^2$G $+$ 35\%  4f$^3$ ($_1^2$G) $^2$G $+$  8\%  4f$^3$ ($_2^2$F) $^2$F $+$  2\%  4f$^3$ ($_1^2$F) $^2$F  \\
    40 &  7/2$-$   &      1.0247182  &    69\%  4f$^3$ ($_1^2$F) $^2$F $+$ 21\%  4f$^3$ ($_2^2$F) $^2$F $+$  4\%  4f$^3$ ($_2^2$G) $^2$G $+$  2\%  4f$^3$ ($_1^2$G) $^2$G $+$  2\%  4f$^3$ ($_1^4$D) $^4$D  \\
    41 &  5/2$-$   &      1.0572781  &    52\%  4f$^3$ ($_1^2$F) $^2$F $+$ 46\%  4f$^3$ ($_2^2$F) $^2$F  \\
   163 & 19/2$+$   &      7.6177821  &    42\%  4d$^9$ 4f$^4$ ($_1^5$I) $^6$K
                                      $+$ 34\%  4d$^9$ 4f$^4$ ($_1^5$I) $^6$L
                                      $+$  6\%  4d$^9$ 4f$^4$ ($_2^3$K) $^4$L \\
       &           &                 & $+$  6\%  4d$^9$ 4f$^4$ ($_2^3$K) $^4$M
                                      $+$  2\%  4d$^9$ 4f$^4$ ($_1^3$K) $^4$M  \\
   172 & 21/2$+$   &      7.6626000  &    56\%  4d$^9$ 4f$^4$ ($_1^5$I) $^6$L $+$ 24\%  4d$^9$ 4f$^4$ ($_2^3$K) $^4$M $+$  5\%  4d$^9$ 4f$^4$ ($_1^3$L) $^4$N  \\
       &           &                 & $+$  5\%  4d$^9$ 4f$^4$ ($_1^3$K) $^4$M $+$  5\%  4d$^9$ 4f$^4$ ($_2^1$L) $^2$N  \\
   199 & 19/2$+$   &      7.7465147  &    20\%  4d$^9$ 4f$^4$ ($_1^5$I) $^6$L
                                      $+$ 18\%  4d$^9$ 4f$^4$ ($_1^3$L) $^4$N
                                      $+$  9\%  4d$^9$ 4f$^4$ ($_1^3$L) $^4$M  \\
       &           &                 & $+$  8\%  4d$^9$ 4f$^4$ ($_1^3$L) $^2$N
                                      $+$  7\%  4d$^9$ 4f$^4$ ($_1^3$K) $^4$M  \\
   204 & 21/2$+$   &      7.7558985  &    27\%  4d$^9$ 4f$^4$ ($_1^5$I) $^6$L $+$ 19\%  4d$^9$ 4f$^4$ ($_1^3$L) $^4$N $+$ 17\%  4d$^9$ 4f$^4$ ($_1^3$M) $^4$O \\ 
       &           &                 & $+$ 14\%  4d$^9$ 4f$^4$ ($_1^3$M) $^2$O $+$ 11\%  4d$^9$ 4f$^4$ ($_2^1$L) $^2$N  \\
   222 & 23/2$+$   &      7.8064391  &    44\%  4d$^9$ 4f$^4$ ($_1^3$M) $^4$O $+$ 42\%  4d$^9$ 4f$^4$ ($_1^3$L) $^4$N $+$  9\%  4d$^9$ 4f$^4$ ($_1^3$M) $^2$O $+$  6\%  4d$^9$ 4f$^4$ ($_1^3$M) $^4$N  \\
   242 & 21/2$+$   &      7.8498415  &    29\%  4d$^9$ 4f$^4$ ($_1^3$L) $^4$N $+$ 23\%  4d$^9$ 4f$^4$ ($_2^1$L) $^2$N $+$ 16\%  4d$^9$ 4f$^4$ ($_1^3$M) $^2$O \\
       &           &                 & $+$ 11\%  4d$^9$ 4f$^4$ ($_1^3$L) $^4$M $+$  6\%  4d$^9$ 4f$^4$ ($_1^3$K) $^4$M  \\
   248 & 25/2$+$   &      7.8577952  &    88\%  4d$^9$ 4f$^4$ ($_1^3$M) $^4$O $+$ 12\%  4d$^9$ 4f$^4$ ($_1^1$N) $^2$Q  \\
   255 & 19/2$+$   &      7.8735085  &    28\%  4d$^9$ 4f$^4$ ($_1^3$I) $^4$L
                                      $+$ 22\%  4d$^9$ 4f$^4$ ($_2^3$K) $^4$L
                                      $+$ 13\%  4d$^9$ 4f$^4$ ($_1^3$L) $^4$N \\
       &           &                 & $+$ 11\%  4d$^9$ 4f$^4$ ($_1^3$L) $^2$N
                                      $+$  6\%  4d$^9$ 4f$^4$ ($_1^5$I) $^6$K  \\
   260 & 21/2$+$   &      7.8851464  &    38\%  4d$^9$ 4f$^4$ ($_2^3$K) $^4$M $+$ 16\%  4d$^9$ 4f$^4$ ($_1^3$M) $^4$O $+$ 14\%  4d$^9$ 4f$^4$ ($_1^5$I) $^6$L \\
       &           &                 & $+$ 13\%  4d$^9$ 4f$^4$ ($_1^3$M) $^2$O $+$ 10\%  4d$^9$ 4f$^4$ ($_2^1$L) $^2$N  \\
   270 & 19/2$+$   &      7.8984569  &    28\%  4d$^9$ 4f$^4$ ($_2^3$K) $^4$M
                                      $+$ 11\%  4d$^9$ 4f$^4$ ($_2^1$L) $^2$M
                                      $+$ 11\%  4d$^9$ 4f$^4$ ($_2^3$K) $^2$M  \\
       &           &                 & $+$  7\%  4d$^9$ 4f$^4$ ($_1^3$M) $^4$N
                                      $+$  6\%  4d$^9$ 4f$^4$ ($_1^1$K) $^2$M  \\
   284 & 23/2$+$   &      7.9247134  &    47\%  4d$^9$ 4f$^4$ ($_1^3$L) $^4$N $+$ 32\%  4d$^9$ 4f$^4$ ($_1^3$M) $^4$O $+$ 17\%  4d$^9$ 4f$^4$ ($_1^3$M) $^2$O $+$  4\%  4d$^9$ 4f$^4$ ($_1^3$M) $^4$N  \\
   297 & 21/2$+$   &      7.9484240  &    26\%  4d$^9$ 4f$^4$ ($_1^3$L) $^4$M
                                      $+$ 19\%  4d$^9$ 4f$^4$ ($_1^3$L) $^2$N
                                      $+$ 14\%  4d$^9$ 4f$^4$ ($_1^3$K) $^4$M  \\
       &           &                 & $+$ 10\%  4d$^9$ 4f$^4$ ($_1^3$M) $^4$M
                                      $+$  9\%  4d$^9$ 4f$^4$ ($_1^3$L) $^4$N  \\
   331 & 19/2$+$   &      8.0053795  &    16\%  4d$^9$ 4f$^4$ ($_2^3$K) $^4$L
                                      $+$ 10\%  4d$^9$ 4f$^4$ ($_1^3$M) $^4$N
                                      $+$ 10\%  4d$^9$ 4f$^4$ ($_1^3$L) $^4$N  \\
       &           &                 & $+$  8\%  4d$^9$ 4f$^4$ ($_2^3$K) $^2$M
                                      $+$  6\%  4d$^9$ 4f$^4$ ($_1^3$M) $^4$M  \\
   348 & 21/2$+$   &      8.0334675  &    36\%  4d$^9$ 4f$^4$ ($_2^1$L) $^2$N $+$ 20\%  4d$^9$ 4f$^4$ ($_1^3$M) $^2$O $+$ 13\%  4d$^9$ 4f$^4$ ($_1^3$M) $^4$N \\ 
       &           &                 & $+$ 13\%  4d$^9$ 4f$^4$ ($_2^3$K) $^4$M $+$  5\%  4d$^9$ 4f$^4$ ($_1^3$K) $^4$M  \\
   350 & 25/2$+$   &      8.0341442  &    88\%  4d$^9$ 4f$^4$ ($_1^1$N) $^2$Q $+$ 12\%  4d$^9$ 4f$^4$ ($_1^3$M) $^4$O  \\
   366 & 23/2$+$   &      8.0573172  &    60\%  4d$^9$ 4f$^4$ ($_1^3$M) $^4$N $+$ 16\%  4d$^9$ 4f$^4$ ($_1^3$M) $^2$O $+$ 11\%  4d$^9$ 4f$^4$ ($_1^3$L) $^4$N \\ 
       &           &                 & $+$  5\%  4d$^9$ 4f$^4$ ($_1^1$N) $^2$O $+$  4\%  4d$^9$ 4f$^4$ ($_1^3$M) $^4$O  \\
   368 & 21/2$+$   &      8.0595059  &    22\%  4d$^9$ 4f$^4$ ($_1^3$M) $^4$N $+$ 15\%  4d$^9$ 4f$^4$ ($_1^3$K) $^4$M $+$ 12\%  4d$^9$ 4f$^4$ ($_2^3$K) $^4$M \\
       &           &                 & $+$ 11\%  4d$^9$ 4f$^4$ ($_1^3$M) $^4$O $+$ 11\%  4d$^9$ 4f$^4$ ($_1^1$L) $^2$N  \\
   392 & 21/2$+$   &      8.0981039  &    38\%  4d$^9$ 4f$^4$ ($_1^3$L) $^4$M $+$ 12\%  4d$^9$ 4f$^4$ ($_1^3$K) $^4$M $+$ 11\%  4d$^9$ 4f$^4$ ($_1^3$M) $^4$N \\
       &           &                 & $+$ 11\%  4d$^9$ 4f$^4$ ($_1^1$L) $^2$N $+$  7\%  4d$^9$ 4f$^4$ ($_2^1$L) $^2$N  \\
   493 & 23/2$+$   &      8.2522591  &    50\%  4d$^9$ 4f$^4$ ($_1^1$N) $^2$O $+$ 40\%  4d$^9$ 4f$^4$ ($_1^1$N) $^2$Q $+$  8\%  4d$^9$ 4f$^4$ ($_1^3$M) $^4$N $+$  1\%  4d$^9$ 4f$^4$ ($_1^3$M) $^2$O  \\
  4130 & 27/2$-$   &     14.9956274  &    33\%  4d$^8$ ($_2^3$F) $^3$F 4f$^5$ ($_0^4$M) $^6$O
                                      $+$ 27\%  4d$^8$ ($_2^3$F) $^3$F 4f$^5$ ($_0^4$M) $^6$Q
                                      $+$ 23\%  4d$^8$ ($_2^3$F) $^3$F 4f$^5$ ($_1^4$L) $^6$O  \\
       &           &                 &     $+$  3\%  4d$^8$ ($_2^1$G) $^1$G 4f$^5$ ($_0^4$M) $^4$Q
                                      $+$  3\%  4d$^8$ ($_2^3$F) $^3$F 4f$^5$ ($_1^2$N) $^4$Q  \\
  4247 & 29/2$-$   &     15.0375603  &    76\%  4d$^8$ ($_2^3$F) $^3$F 4f$^5$ ($_0^4$M) $^6$Q $+$ 14\%  4d$^8$ ($_2^3$F) $^3$F 4f$^5$ ($_1^2$N) $^4$R $+$  7\%  4d$^8$ ($_2^1$G) $^1$G 4f$^5$ ($_0^4$M) $^4$R   \\ 
       &           &                 & $+$ 2\%  4d$^8$ ($_2^1$G) $^1$G 4f$^5$ ($_1^2$N) $^2$T $+$  1\%  4d$^8$ ($_2^3$F) $^3$F 4f$^5$ ($_0^2$O) $^2$T  \\
  4796 & 29/2$-$   &     15.2300106  &    38\%  4d$^8$ ($_2^3$F) $^3$F 4f$^5$ ($_1^2$N) $^4$R $+$ 18\%  4d$^8$ ($_2^3$F) $^3$F 4f$^5$ ($_0^2$O) $^4$T $+$ 14\%  4d$^8$ ($_2^3$F) $^3$F 4f$^5$ ($_0^2$O) $^2$T \\
       &           &                 & $+$ 13\%  4d$^8$ ($_2^3$F) $^3$F 4f$^5$ ($_0^4$M) $^6$Q $+$  7\%  4d$^8$ ($_2^1$G) $^1$G 4f$^5$ ($_1^2$N) $^2$T  \\
  4956 & 31/2$-$   &     15.2822190  &    86\%  4d$^8$ ($_2^3$F) $^3$F 4f$^5$ ($_0^2$O) $^4$T $+$ 14\%  4d$^8$ ($_2^1$G) $^1$G 4f$^5$ ($_0^2$O) $^2$U  \\

\end{longtable}
\end{flushleft}

\endgroup 



\begingroup 

\renewcommand{\arraystretch}{1.2}
\renewcommand{\tabcolsep}{1mm}
\scriptsize

\begin{flushleft}

\LTcapwidth 12cm

\begin{longtable}{ l l l l l l l l l l l l }

\caption{\label{t2}  Configuration interaction strengths $T$ between initial 
$K_{1}$ and admixed $K_{2}$ configurations divided by statistical weight $g_{1}$ 
of initial configuration for some configurations  in $W^{25+}$. Shells of 
admixed configurations  $K_{2}$ marked relatively to the corresponding initial 
configuration $K_{1}$.} \\

\hline
 $K_{1}$ & $T/g_{1}$ & $K_{2}$ && $T/g_{1}$ & $K_{2}$ && $T/g_{1}$ & $K_{2}$ && $T/g_{1}$ & $K_{2}$ \\
\hline
\endfirsthead
\caption[]{ (continued) }  \\
\hline
 $K_{1}$ & $T/g_{1}$ & $K_{2}$ && $T/g_{1}$ & $K_{2}$ && $T/g_{1}$ & $K_{2}$ && $T/g_{1}$ & $K_{2}$ \\
\hline
\endhead
\hline \multicolumn{4}{r}{\textit{Continued on next page}} \\
\endfoot
\hline
\endlastfoot
$4f^{3}$     &   $1.26     {-2}$           &   $4d^{-2}4f^{2}$                  &&   $5.95     {-3}$     &   $4d^{-1}5d^{1}$                  &&   $2.30     {-3}$     &   $4f^{-1}5f^{1}$                  &&   $1.96     {-3}$     &   $4p^{-1}5p^{1}$   \\
            &   $1.52     {-3}$            &   $4d^{-1}5g^{1}$                  &&   $9.23     {-4}$     &   $4p^{-1}4f^{1}$                  &&   $8.06     {-4}$     &   $4d^{-1}6d^{1}$                  &&   $6.61     {-4}$     &   $4d^{-1}6g^{1}$   \\
            &   $4.45     {-4}$            &   $4s^{-1}5s^{1}$                  &&   $4.27     {-4}$     &   $4p^{-2}4f^{2}$                  &&   $3.30     {-4}$     &   $4d^{-1}7g^{1}$                  &&   $3.07     {-4}$     &   $4f^{-1}6f^{1}$   \\
            &   $3.05     {-4}$            &   $4d^{-2}5d^{2}$                  &&   $3.04     {-4}$     &   $4p^{-1}4d^{-1}4f^{1}5g^{1}$     &&   $2.92     {-4}$     &   $4p^{-1}4d^{-1}5p^{1}5d^{1}$     &&   $2.91     {-4}$     &   $4d^{-1}4f^{-1}5d^{1}5f^{1}$   \\
            &   $2.58     {-4}$            &   $4p^{-1}6p^{1}$                  &&   $2.55     {-4}$     &   $4d^{-1}7d^{1}$                  &&   $1.62     {-4}$     &   $4p^{-1}4d^{-1}4f^{1}6g^{1}$     &&   $1.49     {-4}$     &   $4d^{-2}5d^{1}6d^{1}$   \\
$4d^{9} 4f^{4}$     &   $8.48     {-3}$    &   $4d^{-1}5d^{1}$                  &&   $8.41     {-3}$     &   $4d^{-2}4f^{2}$                  &&   $7.05     {-3}$     &   $4d^{1}4f^{-2}5d^{1}$            &&   $6.67     {-3}$     &   $4f^{-1}5f^{1}$   \\
            &   $3.69     {-3}$            &   $4p^{-1}4f^{1}$                  &&   $3.08     {-3}$     &   $4p^{-1}5p^{1}$                  &&   $1.75     {-3}$     &   $4d^{-1}5g^{1}$                  &&   $1.14     {-3}$     &   $4d^{-1}6d^{1}$   \\
            &   $8.86     {-4}$            &   $4f^{-1}6f^{1}$                  &&   $7.62     {-4}$     &   $4d^{-1}6g^{1}$                  &&   $6.99     {-4}$     &   $4s^{-1}5s^{1}$                  &&   $5.33     {-4}$     &   $4d^{1}4f^{-2}5g^{1}$   \\
            &   $4.04     {-4}$            &   $4p^{-1}6p^{1}$                  &&   $3.81     {-4}$     &   $4d^{-1}7g^{1}$                  &&   $3.62     {-4}$     &   $4d^{-1}7d^{1}$                  &&   $3.51     {-4}$     &   $4d^{-1}4f^{-1}5d^{1}5f^{1}$   \\
            &   $3.50     {-4}$            &   $4p^{-2}4f^{2}$                  &&   $2.72     {-4}$     &   $4f^{-1}7f^{1}$                  &&   $2.66     {-4}$     &   $4p^{-1}4d^{-1}5p^{1}5d^{1}$     &&   $2.49     {-4}$     &   $4p^{-1}4d^{-1}4f^{1}5g^{1}$   \\
$4f^{2}5s$     &   $1.52     {-2}$         &   $4d^{-2}4f^{2}$                  &&   $5.63     {-3}$     &   $4d^{-1}4f^{1}5s^{-1}5p^{1}$     &&   $4.43     {-3}$     &   $4d^{-1}5d^{1}$                  &&   $1.49     {-3}$     &   $4p^{-1}5p^{1}$   \\
            &   $1.06     {-3}$            &   $4d^{-1}5g^{1}$                  &&   $7.41     {-4}$     &   $4p^{-1}4f^{1}$                  &&   $5.85     {-4}$     &   $4d^{-1}6d^{1}$                  &&   $5.28     {-4}$     &   $4f^{-1}5f^{1}$   \\
            &   $5.23     {-4}$            &   $5s^{-1}5d^{1}$                  &&   $5.12     {-4}$     &   $4p^{-2}4f^{2}$                  &&   $4.61     {-4}$     &   $4d^{-1}6g^{1}$                  &&   $3.24     {-4}$     &   $4p^{-1}4d^{-1}4f^{1}5g^{1}$   \\
            &   $2.92     {-4}$            &   $4d^{-2}5d^{2}$                  &&   $2.82     {-4}$     &   $4p^{-1}4d^{-1}5p^{1}5d^{1}$     &&   $2.31     {-4}$     &   $4d^{-1}7g^{1}$                  &&   $2.16     {-4}$     &   $4d^{-1}4f^{1}5s^{-1}6p^{1}$   \\
            &   $1.96     {-4}$            &   $4f^{-1}5s^{-1}5p^{1}5g^{1}$     &&   $1.87     {-4}$     &   $4p^{-1}6p^{1}$                  &&   $1.84     {-4}$     &   $4d^{-1}7d^{1}$                  &&   $1.84     {-4}$     &   $4d^{-1}4f^{-1}5d^{1}5f^{1}$   \\
$4f^{2}5p$     &   $1.52     {-2}$         &   $4d^{-2}4f^{2}$                  &&   $5.28     {-3}$     &   $4d^{-1}4f^{1}5s^{1}5p^{-1}$     &&   $4.38     {-3}$     &   $4d^{-1}5d^{1}$                  &&   $2.85     {-3}$     &   $4d^{-1}4f^{1}5p^{-1}5d^{1}$   \\
            &   $1.20     {-3}$            &   $4p^{-1}5p^{1}$                  &&   $1.11     {-3}$     &   $4d^{-1}5g^{1}$                  &&   $8.61     {-4}$     &   $4p^{-1}4f^{1}$                  &&   $5.78     {-4}$     &   $4d^{-1}6d^{1}$   \\
            &   $5.19     {-4}$            &   $4f^{-1}5f^{1}$                  &&   $5.12     {-4}$     &   $4p^{-2}4f^{2}$                  &&   $4.77     {-4}$     &   $4d^{-1}6g^{1}$                  &&   $3.26     {-4}$     &   $4s^{-1}5s^{1}$   \\
            &   $3.23     {-4}$            &   $4p^{-1}4d^{-1}4f^{1}5g^{1}$     &&   $2.91     {-4}$     &   $4d^{-2}5d^{2}$                  &&   $2.38     {-4}$     &   $4d^{-1}7g^{1}$                  &&   $2.34     {-4}$     &   $4p^{-1}4d^{-1}5p^{1}5d^{1}$   \\
            &   $1.96     {-4}$            &   $4p^{-1}6p^{1}$                  &&   $1.84     {-4}$     &   $5p^{-1}5f^{1}$                  &&   $1.83     {-4}$     &   $4d^{-1}4f^{-1}5d^{1}5f^{1}$     &&   $1.83     {-4}$     &   $4d^{-1}7d^{1}$   \\
$4f^{2}5d$     &   $7.76     {-2}$         &   $4d^{-1}4f^{2}5d^{-1}$           &&   $1.52     {-2}$     &   $4d^{-2}4f^{2}$                  &&   $7.10     {-3}$     &   $4d^{-1}4f^{1}5p^{1}5d^{-1}$     &&   $3.86     {-3}$     &   $4d^{-1}5d^{1}$   \\
            &   $2.05     {-3}$            &   $4d^{-1}4f^{1}5d^{-1}5f^{1}$     &&   $1.42     {-3}$     &   $4p^{-1}5p^{1}$                  &&   $1.09     {-3}$     &   $4d^{-1}5g^{1}$                  &&   $8.31     {-4}$     &   $4p^{-1}4f^{1}$   \\
            &   $6.02     {-4}$            &   $4d^{-1}6d^{1}$                  &&   $5.13     {-4}$     &   $4p^{-2}4f^{2}$                  &&   $5.10     {-4}$     &   $4f^{-1}5f^{1}$                  &&   $4.69     {-4}$     &   $4d^{-1}6g^{1}$   \\
            &   $3.51     {-4}$            &   $4f^{-1}5s^{1}5p^{1}5d^{-1}$     &&   $3.23     {-4}$     &   $4p^{-1}4d^{-1}4f^{1}5g^{1}$     &&   $3.22     {-4}$     &   $4s^{-1}5s^{1}$                  &&   $2.52     {-4}$     &   $4p^{-1}4d^{-1}5p^{1}5d^{1}$   \\
            &   $2.34     {-4}$            &   $4d^{-1}7g^{1}$                  &&   $2.32     {-4}$     &   $4d^{-2}5d^{2}$                  &&   $1.93     {-4}$     &   $4d^{-1}7d^{1}$                  &&   $1.85     {-4}$     &   $4p^{-1}6p^{1}$   \\
$4f^{2}5f$     &   $6.14     {-1}$         &   $4d^{-1}4f^{1}5s^{1}5f^{-1}$     &&   $2.38     {-2}$     &   $4f^{-1}5p^{2}5f^{-1}$           &&   $1.52     {-2}$     &   $4d^{-2}4f^{2}$                  &&   $8.16     {-3}$     &   $4d^{-1}4f^{1}5d^{1}5f^{-1}$   \\
            &   $4.27     {-3}$            &   $4d^{-1}5d^{1}$                  &&   $1.41     {-3}$     &   $4p^{-1}5p^{1}$                  &&   $1.10     {-3}$     &   $4d^{-1}5g^{1}$                  &&   $9.73     {-4}$     &   $4d^{-1}4f^{1}5f^{-1}5g^{1}$   \\
            &   $8.02     {-4}$            &   $4p^{-1}4f^{1}$                  &&   $7.22     {-4}$     &   $4f^{-1}5f^{1}$                  &&   $6.58     {-4}$     &   $4f^{1}5f^{-1}$                  &&   $5.70     {-4}$     &   $4d^{-1}6d^{1}$   \\
            &   $5.13     {-4}$            &   $4p^{-2}4f^{2}$                  &&   $4.81     {-4}$     &   $4p^{-1}4f^{1}5p^{1}5f^{-1}$     &&   $4.66     {-4}$     &   $4d^{-1}6g^{1}$                  &&   $3.23     {-4}$     &   $4p^{-1}4d^{-1}4f^{1}5g^{1}$   \\
            &   $3.21     {-4}$            &   $4s^{-1}5s^{1}$                  &&   $2.89     {-4}$     &   $4d^{-2}5d^{2}$                  &&   $2.79     {-4}$     &   $4p^{-1}4d^{-1}5p^{1}5d^{1}$     &&   $2.32     {-4}$     &   $4d^{-1}7g^{1}$   \\
$4f^{2}5g$     &   $1.59     {-1}$         &   $4d^{-1}4f^{1}5p^{1}5g^{-1}$     &&   $1.52     {-2}$     &   $4d^{-2}4f^{2}$                  &&   $4.17     {-3}$     &   $4d^{-1}5d^{1}$                  &&   $3.25     {-3}$     &   $4d^{-1}4f^{2}5g^{-1}$   \\
            &   $2.36     {-3}$            &   $4d^{-1}4f^{1}5f^{1}5g^{-1}$     &&   $1.61     {-3}$     &   $4f^{-1}5p^{1}5d^{1}5g^{-1}$     &&   $1.38     {-3}$     &   $4p^{-1}5p^{1}$                  &&   $1.02     {-3}$     &   $4d^{-1}5g^{1}$   \\
            &   $8.18     {-4}$            &   $4p^{-1}4f^{1}$                  &&   $5.59     {-4}$     &   $4d^{-1}6d^{1}$                  &&   $5.38     {-4}$     &   $4f^{-1}5f^{1}$                  &&   $5.14     {-4}$     &   $4p^{-2}4f^{2}$   \\
            &   $4.69     {-4}$            &   $4d^{-1}6g^{1}$                  &&   $4.65     {-4}$     &   $4d^{-1}4f^{1}5g^{-1}6h1$        &&   $3.13     {-4}$     &   $4s^{-1}5s^{1}$                  &&   $3.04     {-4}$     &   $4p^{-1}4d^{-1}4f^{1}5g^{1}$   \\
            &   $2.88     {-4}$            &   $4d^{-2}5d^{2}$                  &&   $2.77     {-4}$     &   $4p^{-1}4d^{-1}5p^{1}5d^{1}$     &&   $2.33     {-4}$     &   $4d^{-1}7g^{1}$                  &&   $2.01     {-4}$     &   $5g^{-1}6g^{1}$   \\
$4f^{2}6g$     &   $1.52     {-2}$         &   $4d^{-2}4f^{2}$                  &&   $1.28     {-2}$     &   $4p^{-1}4f^{1}5s^{1}6g^{-1}$     &&   $4.06     {-3}$     &   $4d^{-1}5d^{1}$                  &&   $1.36     {-3}$     &   $4p^{-1}5p^{1}$   \\
            &   $1.05     {-3}$            &   $4d^{-1}5g^{1}$                  &&   $1.03     {-3}$     &   $4f^{-1}5s^{1}6p^{1}6g^{-1}$     &&   $7.63     {-4}$     &   $4p^{-1}4f^{1}$                  &&   $6.86     {-4}$     &   $4d^{-1}4f^{2}6g^{-1}$   \\
            &   $5.48     {-4}$            &   $4d^{-1}6d^{1}$                  &&   $5.36     {-4}$     &   $4d^{-1}4f^{1}6f^{1}6g^{-1}$     &&   $5.13     {-4}$     &   $4p^{-2}4f^{2}$                  &&   $4.63     {-4}$     &   $4f^{-1}5f^{1}$   \\
            &   $4.26     {-4}$            &   $4d^{-1}6g^{1}$                  &&   $3.20     {-4}$     &   $4p^{-1}4d^{-1}4f^{1}5g^{1}$     &&   $3.10     {-4}$     &   $4s^{-1}5s^{1}$                  &&   $2.85     {-4}$     &   $4d^{-2}5d^{2}$   \\
            &   $2.75     {-4}$            &   $4p^{-1}4d^{-1}5p^{1}5d^{1}$     &&   $2.26     {-4}$     &   $4d^{-1}7g^{1}$                  &&   $2.01     {-4}$     &   $5g^{1}6g^{-1}$                  &&   $1.80     {-4}$     &   $6g^{-1}7g^{1}$   \\
\end{longtable}

\end{flushleft}

\endgroup


\begin{figure}
 \includegraphics[scale=0.41]{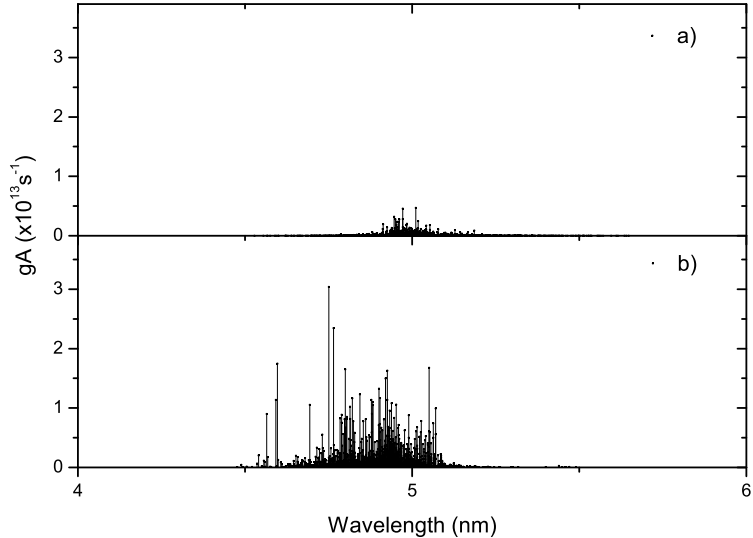}%
 \caption{\label{ci_4f25d_4f3} Calculated transition data obtained using 
a) single-configuration and b) configuration mixing methods for 
$4f^{2}5d \rightarrow 4f^{3}$ transition in $W^{25+}$. }
\end{figure}

Table \ref{t2} presents configuration interaction strengths \cite{Karazija1, 
1997ps_55_667_kucas} for $4f^{3}$, $4d^{9}4f^{4}$, $4f^{2}5s$, $4f^{2}5p$, 
$4f^{2}5d$, $4f^{2}5f$, $4f^{2}5g$, and $4f^{2}6g$ configurations, which are 
responsible for the line formation in emission spectra of 2--30 nm range 
\cite{2010jpb_43_205004_harte, 2011jpb_44_175004_suzuki, 2012jpb_45_205002_harte}. 
The largest configuration mixing is determined for $4f^{2}5d$, $4f^{2}5f$, and 
$4f^{2}5g$ configurations (Table \ref{t2}). Previous analysis has established 
that the correlation effects make a small contribution to the $gA$ spectrum of 
the $W^{25+}$ ion \cite{2010jpb_43_205004_harte}. However, our calculations show 
that the CI increases the radiative transition probabilities for the
$4f^{2}5d \rightarrow 4f^{3}$ transition by an order (Fig. \ref{ci_4f25d_4f3}) 
and  affects  the distribution of lines. The inclusion of correlation effects 
changes the average wavelength and width of lines  ($\bar{\lambda}=5.02$ nm, 
$\sigma=0.54$ nm to $\bar{\lambda}=4.90$ nm, $\sigma=1.20$ nm). Thus, the 
spectra from these transitions in $W^{25+}$ have to be investigated by taking 
into account the configuration mixing. It is interesting to note that the 
strongest mixing of $4d^{9}4f^{4}$ configuration does not take place with 
$4f^{2}5d$ configuration. The configuration interaction strength shows that 
$4d^{8}4f^{4}5d$ and $4d^{7}4f^{6}$ configurations have the largest influence 
to the $4d^{9}4f^{4}$ configuration. The additional CI basis has been used to 
estimate the impact of these configurations ($4d^{8}4f^{4}5d$, $4d^{7}4f^{6}$, 
$4p^{5}4d^{9}4f^{5}$) on the wavelengths and radiative transition probabilities 
of $4d^{9}4f^{4} \rightarrow 4f^{3}$ transition. It was found that the 
transition wavelengths increase by less than 0.17 nm while the radiative 
transition probabilities change by less than 20 \% for the strongest lines.


\begingroup

\begin{flushleft}

\renewcommand{\arraystretch}{1.2}
\renewcommand{\tabcolsep}{1mm}
\scriptsize

\LTcapwidth 12cm

\begin{longtable}{ r r l r l r l r l r l r }

\caption{\label{t3} The five major spontaneous radiative transition probabilities 
$A^r$ (in $s^{-1}$) from each level having the largest lifetimes are presented. 
Arrow marks the final level to which radiative transition happens from the level. 
The sum of all radiative probabilities from the corresponding level is given in 
the last column. $a\pm b = a\times 10^{\pm b}$}. \\
\hline\hline
 Initial & $A^r$ ($s^{-1}$) & final & $A^r$ ($s^{-1}$) & final & $A^r$ ($s^{-1}$) & final & $A^r$ ($s^{-1}$) & final & $A^r$ ($s^{-1}$) & final & $\sum$ $A^r$ ($s^{-1}$) \\ 
 level &&level&&level&&level&&level&&level& \\ 
\hline
\endfirsthead
\caption[]{ (continued) }  \\
\hline
 Initial & $A^r$ ($s^{-1}$) & final & $A^r$ ($s^{-1}$) & final & $A^r$ ($s^{-1}$) & final & $A^r$ ($s^{-1}$) & final & $A^r$ ($s^{-1}$) & final & $\sum$ $A^r$ ($s^{-1}$) \\ 
 level &&level&&level&&level&&level&&level& \\ 
\hline
\endhead
\hline \multicolumn{4}{r}{\textit{Continued on next page}} \\
\endfoot
\hline
\endlastfoot
        2 & 2.908$+$2 & $\rightarrow$    1 &           &      &           &      &           &      &           &       & 2.908$+$2 \\
        4 & 2.125$+$2 & $\rightarrow$    2 & 3.613$-$4 & $\rightarrow$    1 &           &      &           &      &           &       & 2.125$+$2 \\
        5 & 3.552$+$2 & $\rightarrow$    1 & 5.387$+$1 & $\rightarrow$    2 & 3.015$-$6 & $\rightarrow$    4 &           &      &           &       & 4.091$+$2 \\
        6 & 6.176$+$1 & $\rightarrow$    3 & 6.280$-$2 & $\rightarrow$    1 & 1.685$-$7 & $\rightarrow$    5 &           &      &           &       & 6.182$+$1 \\
        7 & 9.194$+$1 & $\rightarrow$    4 & 7.306$-$5 & $\rightarrow$    2 &           &      &           &      &           &       & 9.194$+$1 \\
        8 & 7.810$+$1 & $\rightarrow$    3 & 1.021$+$0 & $\rightarrow$    6 &           &      &           &      &           &       & 7.912$+$1 \\
        9 & 3.703$+$1 & $\rightarrow$    1 & 1.324$+$1 & $\rightarrow$    5 & 2.801$+$0 & $\rightarrow$    6 & 3.169$-$3 & $\rightarrow$    2 & 9.071$-$6 & $\rightarrow$    3  & 5.308$+$1 \\
       10 & 3.954$+$1 & $\rightarrow$    3 & 2.493$+$0 & $\rightarrow$    6 & 8.852$-$1 & $\rightarrow$    1 & 2.228$-$1 & $\rightarrow$    9 & 1.633$-$2 & $\rightarrow$    8  & 4.316$+$1 \\
       11 & 1.114$+$2 & $\rightarrow$    6 & 2.820$+$1 & $\rightarrow$    1 & 1.375$+$1 & $\rightarrow$    5 & 1.048$+$1 & $\rightarrow$   10 & 8.532$+$0 & $\rightarrow$    9  & 1.726$+$2 \\
       12 & 2.338$+$2 & $\rightarrow$    5 & 7.144$+$1 & $\rightarrow$    2 & 5.331$+$1 & $\rightarrow$    1 & 3.649$+$1 & $\rightarrow$    9 & 2.009$-$1 & $\rightarrow$   11  & 3.952$+$2 \\
       13 & 1.813$+$2 & $\rightarrow$    5 & 1.133$+$2 & $\rightarrow$    2 & 1.065$+$2 & $\rightarrow$    4 & 1.732$+$0 & $\rightarrow$    1 & 4.639$-$2 & $\rightarrow$   12  & 4.029$+$2 \\
       14 & 4.712$+$2 & $\rightarrow$    2 & 1.109$+$2 & $\rightarrow$    4 & 6.329$-$2 & $\rightarrow$    1 & 1.203$-$3 & $\rightarrow$    7 & 1.066$-$3 & $\rightarrow$    5  & 5.822$+$2 \\
       15 & 3.031$+$2 & $\rightarrow$    8 & 1.648$+$2 & $\rightarrow$    6 & 8.315$+$1 & $\rightarrow$    3 & 3.671$+$0 & $\rightarrow$   10 & 2.564$-$3 & $\rightarrow$    9  & 5.547$+$2 \\
       16 & 2.139$+$2 & $\rightarrow$   10 & 1.166$+$2 & $\rightarrow$    9 & 9.157$+$1 & $\rightarrow$    6 & 4.024$+$0 & $\rightarrow$    5 & 1.756$+$0 & $\rightarrow$   11  & 4.312$+$2 \\
       17 & 1.944$+$2 & $\rightarrow$    3 & 1.018$+$2 & $\rightarrow$    8 & 4.165$-$1 & $\rightarrow$   15 & 1.112$-$2 & $\rightarrow$   10 & 2.230$-$3 & $\rightarrow$    6  & 2.966$+$2 \\
       18 & 1.360$+$2 & $\rightarrow$    5 & 1.334$+$2 & $\rightarrow$   11 & 6.105$+$1 & $\rightarrow$    2 & 4.338$+$1 & $\rightarrow$    9 & 4.797$+$0 & $\rightarrow$   12  & 3.844$+$2 \\
       19 & 2.822$+$2 & $\rightarrow$    4 & 2.359$+$2 & $\rightarrow$    7 & 2.083$+$1 & $\rightarrow$   14 & 1.528$-$3 & $\rightarrow$    2 & 3.645$-$5 & $\rightarrow$   13  & 5.389$+$2 \\
       20 & 1.318$+$2 & $\rightarrow$    3 & 5.609$+$1 & $\rightarrow$   10 & 4.214$+$1 & $\rightarrow$    6 & 3.839$+$1 & $\rightarrow$   11 & 1.694$+$1 & $\rightarrow$    9  & 2.920$+$2 \\
       21 & 5.476$+$1 & $\rightarrow$    4 & 1.777$+$1 & $\rightarrow$   18 & 1.767$+$1 & $\rightarrow$    5 & 1.605$+$1 & $\rightarrow$    1 & 1.489$+$1 & $\rightarrow$   13  & 1.365$+$2 \\
       22 & 1.558$+$2 & $\rightarrow$    8 & 2.554$+$1 & $\rightarrow$   10 & 7.869$+$0 & $\rightarrow$   20 & 7.033$+$0 & $\rightarrow$   17 & 2.628$+$0 & $\rightarrow$    3  & 2.010$+$2 \\
       23 & 2.175$+$2 & $\rightarrow$    9 & 9.525$+$1 & $\rightarrow$    8 & 6.324$+$1 & $\rightarrow$   15 & 5.260$+$1 & $\rightarrow$   10 & 2.712$+$1 & $\rightarrow$    6  & 4.905$+$2 \\
       24 & 3.813$+$2 & $\rightarrow$   12 & 1.119$+$2 & $\rightarrow$   16 & 9.024$+$1 & $\rightarrow$   11 & 5.966$+$1 & $\rightarrow$   13 & 2.964$+$1 & $\rightarrow$    2  & 7.139$+$2 \\
       25 & 1.737$+$2 & $\rightarrow$   13 & 8.124$+$1 & $\rightarrow$   18 & 6.860$+$1 & $\rightarrow$    5 & 2.709$+$1 & $\rightarrow$    1 & 1.325$+$1 & $\rightarrow$   21  & 3.826$+$2 \\
       26 & 8.403$+$1 & $\rightarrow$   15 & 6.532$+$1 & $\rightarrow$   22 & 5.031$+$1 & $\rightarrow$    8 & 3.473$+$1 & $\rightarrow$   17 & 4.625$-$1 & $\rightarrow$   10  & 2.353$+$2 \\
       27 & 4.311$+$2 & $\rightarrow$   15 & 2.191$+$2 & $\rightarrow$   17 & 2.177$+$2 & $\rightarrow$    6 & 5.601$+$1 & $\rightarrow$   20 & 2.948$+$1 & $\rightarrow$   10  & 9.938$+$2 \\
       28 & 9.272$+$1 & $\rightarrow$   14 & 7.731$+$1 & $\rightarrow$    7 & 3.507$+$1 & $\rightarrow$   13 & 2.866$+$1 & $\rightarrow$    4 & 1.835$+$1 & $\rightarrow$   21  & 2.870$+$2 \\
       29 & 1.584$+$2 & $\rightarrow$   20 & 3.727$+$1 & $\rightarrow$   23 & 1.311$+$1 & $\rightarrow$   16 & 1.252$+$1 & $\rightarrow$   18 & 1.088$+$1 & $\rightarrow$   11  & 2.538$+$2 \\
       30 & 3.985$+$2 & $\rightarrow$   14 & 2.121$+$2 & $\rightarrow$   19 & 6.305$+$1 & $\rightarrow$    4 & 2.850$+$1 & $\rightarrow$    7 & 5.054$-$2 & $\rightarrow$    2  & 7.022$+$2 \\
       31 & 2.155$+$2 & $\rightarrow$   16 & 9.201$+$1 & $\rightarrow$   11 & 8.567$+$1 & $\rightarrow$   18 & 7.530$+$1 & $\rightarrow$   13 & 2.629$+$1 & $\rightarrow$   21  & 5.651$+$2 \\
       32 & 1.863$+$2 & $\rightarrow$   19 & 5.689$+$1 & $\rightarrow$    7 & 7.065$-$1 & $\rightarrow$   30 & 2.982$-$2 & $\rightarrow$    4 & 6.163$-$3 & $\rightarrow$   14  & 2.439$+$2 \\
       33 & 3.112$+$2 & $\rightarrow$   11 & 2.763$+$2 & $\rightarrow$   20 & 2.397$+$2 & $\rightarrow$   10 & 1.311$+$2 & $\rightarrow$   22 & 9.485$+$1 & $\rightarrow$   15  & 1.376$+$3 \\
       34 & 3.113$+$2 & $\rightarrow$   23 & 2.414$+$2 & $\rightarrow$   22 & 1.260$+$2 & $\rightarrow$   26 & 1.070$+$2 & $\rightarrow$   15 & 7.341$+$1 & $\rightarrow$   17  & 8.973$+$2 \\
       35 & 3.340$+$2 & $\rightarrow$   25 & 2.618$+$2 & $\rightarrow$   18 & 6.399$+$1 & $\rightarrow$   12 & 5.837$+$1 & $\rightarrow$   14 & 5.116$+$1 & $\rightarrow$   24  & 9.161$+$2 \\
       36 & 3.681$+$2 & $\rightarrow$   18 & 2.599$+$2 & $\rightarrow$   12 & 1.387$+$2 & $\rightarrow$   16 & 1.097$+$2 & $\rightarrow$   33 & 8.845$+$1 & $\rightarrow$   29  & 1.102$+$3 \\
       37 & 3.591$+$2 & $\rightarrow$   11 & 2.570$+$2 & $\rightarrow$   27 & 2.377$+$2 & $\rightarrow$   29 & 1.013$+$2 & $\rightarrow$   33 & 9.612$+$1 & $\rightarrow$   10  & 1.304$+$3 \\
       38 & 1.856$+$2 & $\rightarrow$   21 & 1.109$+$2 & $\rightarrow$   31 & 9.051$+$1 & $\rightarrow$   24 & 8.616$+$1 & $\rightarrow$   11 & 2.971$+$1 & $\rightarrow$   13  & 5.600$+$2 \\
       39 & 1.010$+$2 & $\rightarrow$   31 & 9.077$+$1 & $\rightarrow$    6 & 6.030$+$1 & $\rightarrow$   10 & 5.205$+$1 & $\rightarrow$   20 & 3.895$+$1 & $\rightarrow$   29  & 4.116$+$2 \\
       40 & 6.083$+$2 & $\rightarrow$   33 & 5.372$+$2 & $\rightarrow$   29 & 3.206$+$2 & $\rightarrow$   39 & 2.216$+$2 & $\rightarrow$   38 & 1.390$+$2 & $\rightarrow$   18  & 2.291$+$3 \\
       41 & 3.458$+$2 & $\rightarrow$   39 & 3.151$+$2 & $\rightarrow$   36 & 1.407$+$2 & $\rightarrow$   20 & 1.015$+$2 & $\rightarrow$   33 & 9.812$+$1 & $\rightarrow$   27  & 1.461$+$3 \\
      172 & 1.814$+$1 & $\rightarrow$  163 & 5.520$+$0 & $\rightarrow$   32 & 3.331$+$0 & $\rightarrow$    7 & 9.091$-$1 & $\rightarrow$   19 & 2.132$-$2 & $\rightarrow$   30  & 2.792$+$1 \\
      204 & 1.451$+$2 & $\rightarrow$  172 & 4.946$+$1 & $\rightarrow$  163 & 1.353$+$1 & $\rightarrow$   32 & 1.029$+$1 & $\rightarrow$   19 & 1.787$+$0 & $\rightarrow$   30  & 2.216$+$2 \\
      222 & 8.664$+$1 & $\rightarrow$  172 & 2.509$+$1 & $\rightarrow$  204 & 7.509$+$0 & $\rightarrow$   32 & 3.216$-$4 & $\rightarrow$  163 & 2.847$-$6 & $\rightarrow$  199  & 1.192$+$2 \\
      242 & 2.567$+$2 & $\rightarrow$   32 & 9.733$+$1 & $\rightarrow$  199 & 4.496$+$1 & $\rightarrow$  163 & 3.558$+$1 & $\rightarrow$  204 & 1.602$+$1 & $\rightarrow$   19  & 5.005$+$2 \\
      248 & 1.982$+$1 & $\rightarrow$  222 & 3.057$-$4 & $\rightarrow$  172 & 6.850$-$6 & $\rightarrow$  204 & 3.217E$-$10 & $\rightarrow$  242 &           &       & 1.982$+$1 \\
      260 & 3.697$+$2 & $\rightarrow$  204 & 9.626$+$1 & $\rightarrow$  163 & 7.476$+$1 & $\rightarrow$  172 & 6.964$+$1 & $\rightarrow$  199 & 4.832$+$1 & $\rightarrow$  222  & 7.065$+$2 \\
      284 & 1.021$+$2 & $\rightarrow$  222 & 6.843$+$1 & $\rightarrow$  242 & 3.351$+$1 & $\rightarrow$  248 & 1.895$+$1 & $\rightarrow$  172 & 1.101$+$1 & $\rightarrow$   32  & 2.387$+$2 \\
      348 & 3.282$+$2 & $\rightarrow$  260 & 2.211$+$2 & $\rightarrow$  270 & 1.363$+$2 & $\rightarrow$  242 & 6.810$+$1 & $\rightarrow$  284 & 5.298$+$1 & $\rightarrow$  255  & 8.931$+$2 \\
      350 & 2.059$+$2 & $\rightarrow$  222 & 1.724$+$2 & $\rightarrow$  248 & 2.040$+$1 & $\rightarrow$  284 & 5.335$-$3 & $\rightarrow$  204 & 1.537$-$3 & $\rightarrow$  172  & 3.987$+$2 \\
      366 & 1.611$+$2 & $\rightarrow$  248 & 5.406$+$1 & $\rightarrow$  297 & 1.384$+$1 & $\rightarrow$  260 & 1.298$+$1 & $\rightarrow$  172 & 5.480$+$0 & $\rightarrow$  284  & 2.516$+$2 \\
      368 & 1.025$+$3 & $\rightarrow$   32 & 2.482$+$2 & $\rightarrow$  222 & 8.422$+$1 & $\rightarrow$  260 & 5.449$+$1 & $\rightarrow$  284 & 5.195$+$1 & $\rightarrow$    7  & 1.674$+$3 \\
      392 & 7.199$+$2 & $\rightarrow$   32 & 2.098$+$2 & $\rightarrow$    7 & 1.274$+$2 & $\rightarrow$  222 & 1.071$+$2 & $\rightarrow$  260 & 6.180$+$1 & $\rightarrow$  331  & 1.613$+$3 \\
      493 & 5.645$+$2 & $\rightarrow$  350 & 3.231$+$2 & $\rightarrow$  248 & 2.727$+$2 & $\rightarrow$  366 & 1.379$+$2 & $\rightarrow$   32 & 8.289$+$1 & $\rightarrow$  368  & 1.625$+$3 \\
     4247 & 1.734$+$3 & $\rightarrow$  248 & 1.212$+$1 & $\rightarrow$ 4130 & 3.621$+$0 & $\rightarrow$  350 & 2.497$+$0 & $\rightarrow$  366 & 1.638$+$0 & $\rightarrow$  222  & 1.755$+$3 \\
     4796 & 9.590$+$2 & $\rightarrow$  248 & 8.271$+$2 & $\rightarrow$  350 & 4.071$+$2 & $\rightarrow$ 4247 & 9.289$+$1 & $\rightarrow$ 4130 & 7.794$+$1 & $\rightarrow$  222  & 2.397$+$3 \\
     4956 & 1.340$+$2 & $\rightarrow$  248 & 3.454$+$1 & $\rightarrow$ 4247 & 1.098$+$1 & $\rightarrow$ 4796 & 9.944$-$1 & $\rightarrow$  350 & 3.145$-$4 & $\rightarrow$ 4130  & 1.805$+$2 \\
\end{longtable}

\end{flushleft}

\endgroup

Electron-impact excitations from the long lived levels are responsible for the 
population of levels in a low density plasma. Table \ref{t3} presents levels, 
which have the smallest values of the total radiative decay rates. We only 
present levels with the total radiative transition rates not exceeding the total 
decay rates of  the ground configuration levels. It should be noted that the 
second excited level can decay to the ground level only through the magnetic 
octupole transition ($\lambda = 2790.29$ nm) which has not been studied here. 
The highly-excited states of the ion have radiative lifetimes comparable with 
the lifetimes of the excited levels of the ground configuration which decay 
mainly through the weak magnetic dipole transitions. These levels belong to the  
$4d^{9}4f^{4}$ and $4d^{8}4f^{5}$ configurations. The extremely large total 
angular momentum $J$ values of these levels limit the possible decay channels 
and lead to the large lifetimes.

The previous investigations \cite{2010jpb_43_205004_harte, 
2011jpb_44_175004_suzuki, 2012jpb_45_205002_harte} show that the lines from the 
electric dipole $(n=4) \rightarrow (n=4)$, $(n=5) \rightarrow (n=4)$, and 
$(n=5) \rightarrow (n=5)$ transitions  in $W^{25+}$ ion have wavelengths in the 
2 -- 30 nm region. The lines in the range from 2 to 4 nm have been identified as 
corresponding to $4f^{N-1}5g \rightarrow 4f^{N}$ ($W^{20+}$--$W^{27+}$) and 
$4f^{N-2}5s5g \rightarrow 4f^{N-1}5s$ ($W^{20+}$--$W^{26+}$) transitions 
\cite{2010jpb_43_205004_harte, 2012jpb_45_205002_harte}. The study 
\cite{2012jpb_45_205002_harte} has included $4f^{N-2}5s5g \rightarrow 4f^{N-1}5s$ 
transition because the first excited configurations of the considered ions 
cannot combine radiatively with the ground configuration. However, correlation 
effects open the decay channels for electric dipole transitions as it is 
demonstrated above for the $W^{25+}$ ion. Furthermore, the levels of $4f^{2}5s$ 
configuration do not have the largest lifetimes among the investigated levels. 


\begin{figure}
 \includegraphics[scale=0.5]{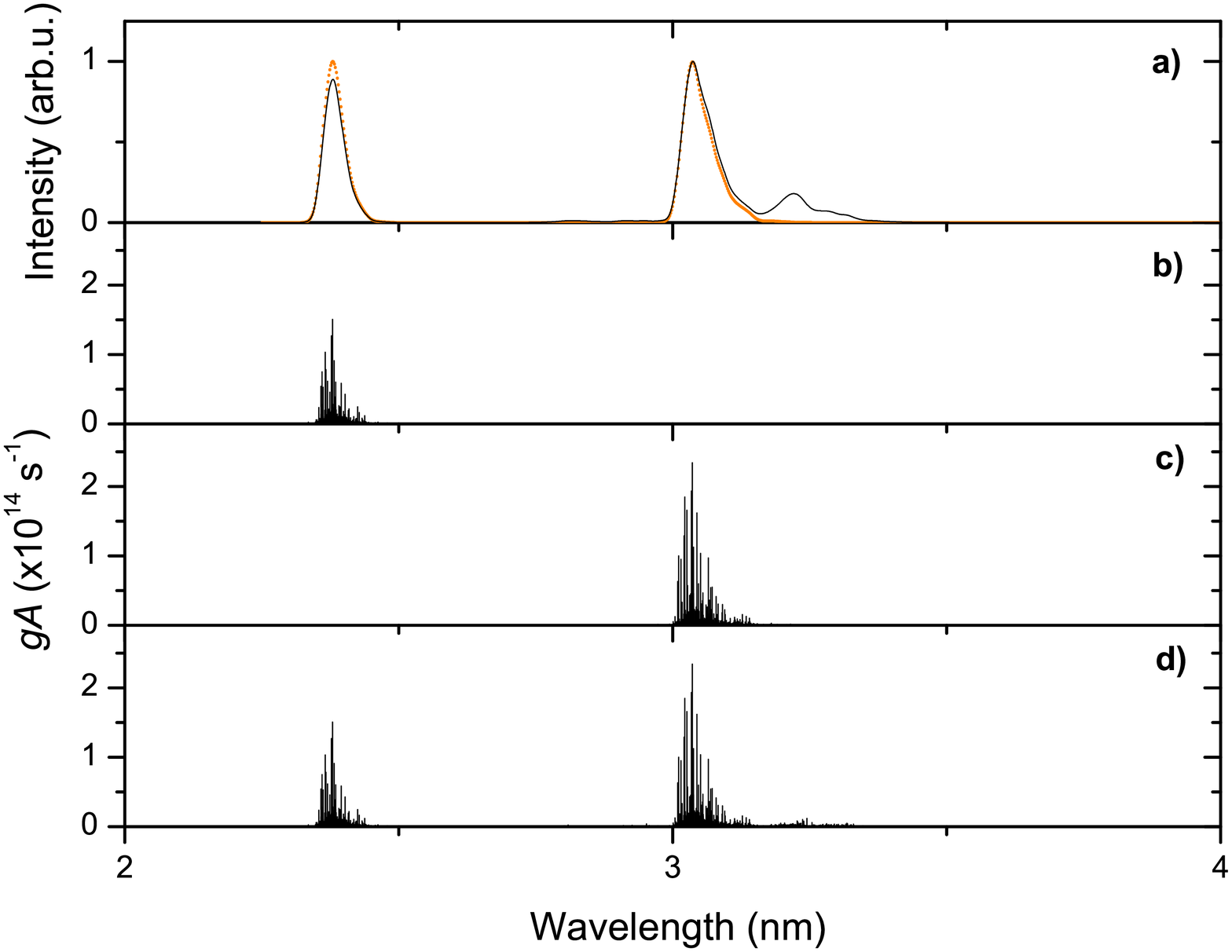}%
 \caption{\label{w25_2_4} Calculated transition data for $W^{25+}$ in the 
2 -- 4 nm range. a) Convoluted $gA$ spectra with a full width at half maximum  
of 0.02 nm. Orange line corresponds to the summed contribution from b) 
$4f^{2}6g \rightarrow 4f^{3}$ and c) $4f^{2}5g \rightarrow 4f^{3}$ transitions. 
Black line comes from d) spectrum of all transitions considered in this work.
}
\end{figure}


\begin{figure}
 \includegraphics[scale=0.33]{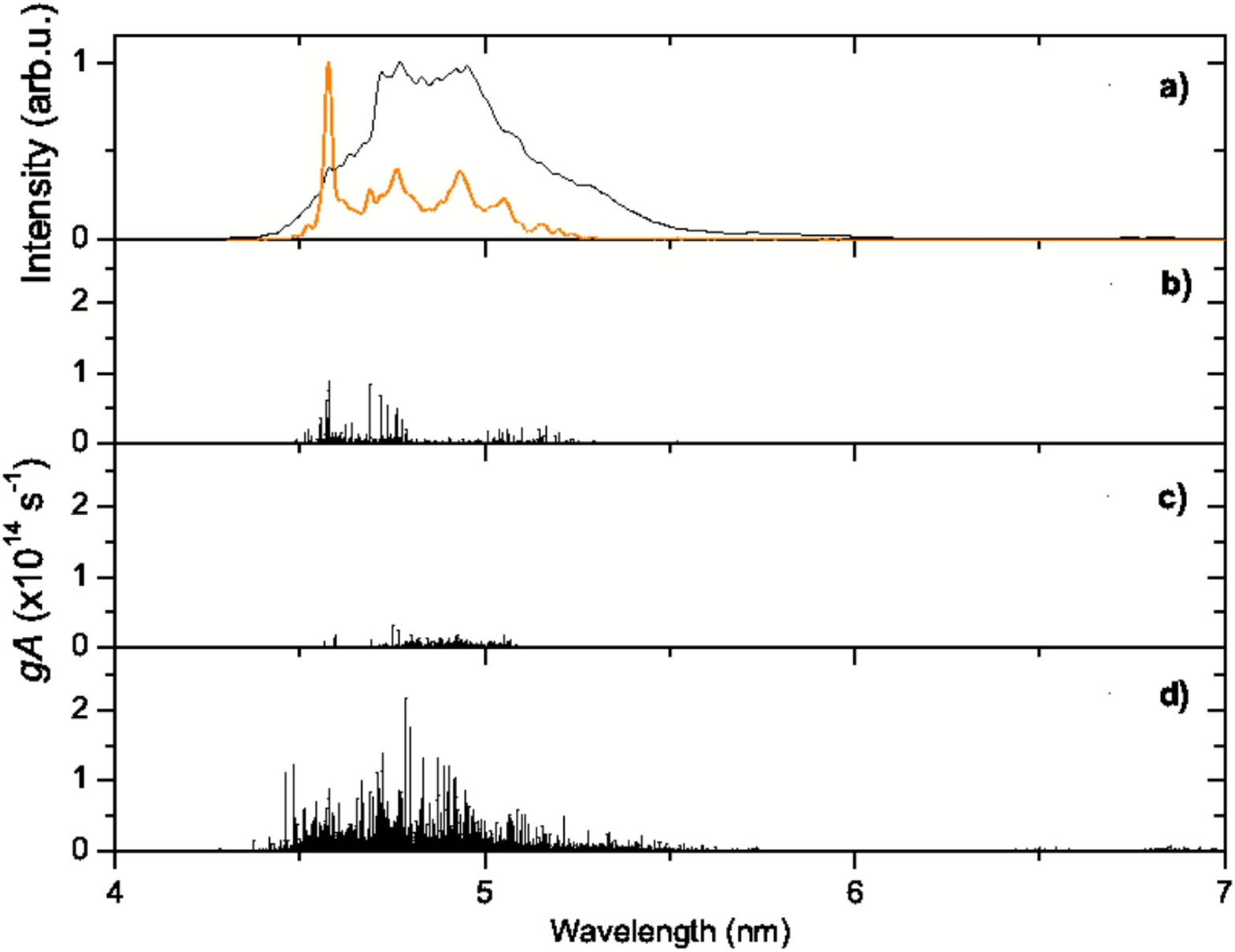}%
 \caption{\label{w25_4_7}  Calculated transition data for $W^{25+}$ in the 
4 -- 7 nm range. a) Convoluted $gA$ spectra with a full width at half maximum  
of 0.02 nm. Orange line corresponds to the summed contribution from 
b) $4d^{9}4f^{4} \rightarrow 4f^{3}$ and c) $4f^{2}5d \rightarrow 4f^{3}$ 
transitions. Black line comes from d) spectrum of all transitions considered 
in this work. 
}
\end{figure}


\begin{figure}
 \includegraphics[scale=0.48]{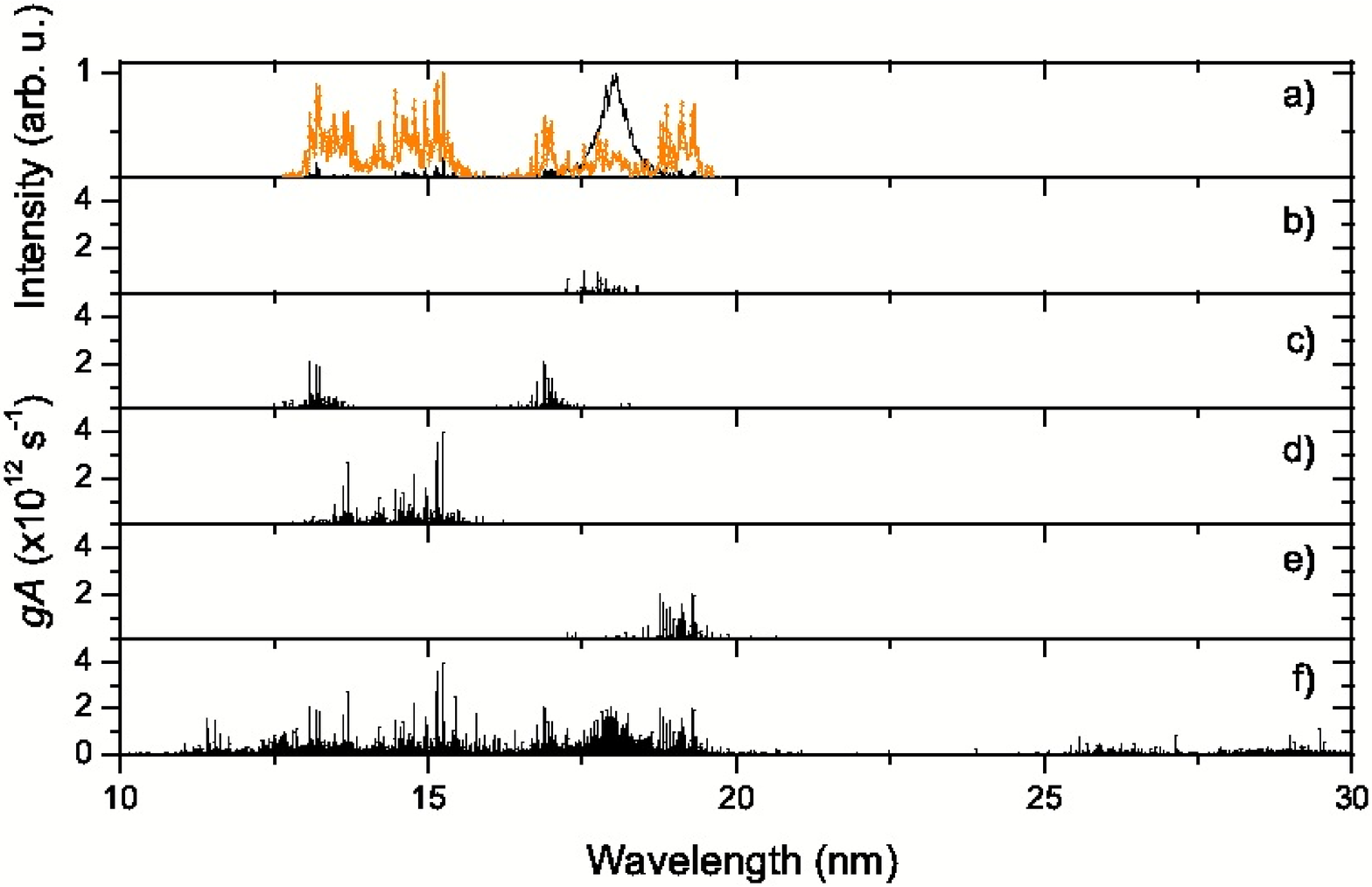}%
 \caption{\label{w25_10_30} Calculated transition data for $W^{25+}$ in the 
10 -- 30 nm range. a) Convoluted $gA$ spectra with a full width at half maximum  
of 0.02 nm. Orange line corresponds to the summed contribution from b) 
$4f^{2}5p \rightarrow 4f^{2}5s$; c) $4f^{2}5d \rightarrow 4f^{2}5p$; d) 
$4f^{2}5f \rightarrow 4f^{2}5d$, and e) $4f^{2}5g \rightarrow 4f^{2}5f$ 
transitions. Black line comes from f) spectrum of all transitions considered in 
this work. 
}
\end{figure}

Figures \ref{w25_2_4}, \ref{w25_4_7}, and \ref{w25_10_30} show theoretical 
radiative transition probabilities multiplied by the statistical weights of 
initial levels. The strongest lines in the spectra are concentrated at 2.4, 3.0,  
and 5.0 nm. The first two groups of the lines correspond mainly to 
$4f^{2}6g \rightarrow 4f^{3}$ and $4f^{2}5g \rightarrow 4f^{3}$ transitions. 
The additional peaks at 3.22 nm arise from the electric dipole 
$4d^{9}4f^{3}5p \rightarrow 4f^{3}$ transitions. Those lines from 
$4f^{2}5g \rightarrow 4f^{3}$  and $4d^{9}4f^{3}5p \rightarrow 4f^{3}$  
transitions have been already identified in large helical device plasmas with 
the wavelengths $\lambda=3.09$ nm and $\lambda=3.25$ nm, respectively 
\cite{2007pfr_2_s1060_chowdhuri}. The identification is based on the data from 
\cite{1988pla_127_255_Finkenthal} where the Unresolved Transition Array (UTA) 
approximation is used in calculations (3.09, 3.24 nm). However, the UTA 
calculations provide  the mean wavelengths slightly higher compared to the peak 
values due to the asymetric nature of $gA$ spectra of these transitions. The UTA 
mean wavelengths calculated using the pseudorelativistic approach 
\cite{2010jpb_43_205004_harte} provides similar values (3.11, 3.22 nm) in 
comparison to the experimental wavelengths of corresponding peaks. 
Our wavelengths (3.04, 3.22 nm) are shorter than the experimental ones. On the 
other hand, $gA$ values for $4d^{9}4f^{3}5p \rightarrow 4f^{3}$  transition are 
about six times smaller than those for $4f^{2}5g \rightarrow 4f^{3}$ transition. 
Therefore, it is unlikely that $4d^{9}4f^{3}5p \rightarrow 4f^{3}$ transition 
would influence the line formation in the observed spectra. Lines corresponding 
to $4f^{2}5g \rightarrow 4f^{3}$ transition have been also observed in  
\cite{2011cjp_89_591_podpaly} but their observed  wavelength $\lambda=3.10$ nm 
is slightly longer compared with the previous value of $3.09$ nm 
\cite{2007pfr_2_s1060_chowdhuri}. On the other hand, our calculations show that 
one of their unidentified lines with $\lambda=2.41$ nm 
\cite{2011cjp_89_591_podpaly} belongs to the $4f^{2}6g \rightarrow 4f^{3}$ 
transition for which $gA$ values have a peak at $2.38$ nm. This transition is 
not considered in \cite{2012jpb_45_205002_harte} assuming that high $n$, $l$ 
states would be collisionally destroyed in a dense laser produced plasma. 

The third group of lines at 5.0 nm originates from $4d^{9}4f^{4} \rightarrow 4f^{3}$ 
and $4f^{2}5d \rightarrow 4f^{3}$ transitions with the  large contribution from 
$4d^{8}4f^{5} \rightarrow 4d^{9}4f^{4}$ transition (Fig. \ref{w25_4_7}). 
From the energy level spectrum (Fig. \ref{w25energy}) one can see that 
$4f^{2}5d$ and $4d^{9}4f^{4}$ configurations overlap. The width of $4f^{2}5d$ 
configuration is about few times smaller than the width of  $4d^{9}4f^{4}$ 
configuration. Since wavelengths of the strongest lines from 
$4d^{9}4f^{4} \rightarrow 4f^{3}$, and $4f^{2}5d \rightarrow 4f^{3}$ transitions 
concentrate in the same area (Fig. \ref{w25_4_7}b and \ref{w25_4_7}c), it 
suggests that the transitions take place from the same energetic region.  

Region of 10--30 nm is covered by lines which originate from 
$4f^{2}5g \rightarrow 4f^{2}5f$, $4f^{2}5f \rightarrow 4f^{2}5d$, 
$4f^{2}5d \rightarrow 4f^{2}5p$, $4f^{2}5p \rightarrow 4f^{2}5s$, 
$4d^{9}4f^{3}5p \rightarrow 4d^{9}4f^{3}5s$, $4f^{2}6d \rightarrow 4f^{2}5f$, 
and $4f5p5d \rightarrow 4f5p^{2}$ transitions in $W^{25+}$ 
(Fig. \ref{w25_10_30}). The obtained results agree well with the other 
calculations \cite{2011jpb_44_175004_suzuki}. Large transition probabilities for 
the lines near 18 nm correspond mainly to 
$4d^{9}4f^{3}5p \rightarrow 4d^{9}4f^{3}5s$ transitions.

\section{The modeling of emission spectra}

Many strong transitions fall into the range of 2 -- 30 nm where strong emission 
from the considered configurations is observed (Figs. \ref{w25_2_4}, 
\ref{w25_4_7}, and \ref{w25_10_30}). Intensive lines in the low-density plasma  
mainly originate from the configurations which correspond to the one-electron 
excitations from the configuration levels with the large lifetimes because such 
levels are mainly populated. We have calculated the electron-impact excitation 
cross-sections from the ground $4f_{5/2}^3$ subconfiguration in the 
distorted-wave approximation at 790 eV electron beam energy, which corresponds 
to the energy used in EBIT measurements \cite{Radtke2001PhysRevA_64_012720}. 
The single-configuration approximation and the UTA mode implemented in Flexible 
Atomic Code (FAC) \cite{2008cjp_86_675_Gu} were applied. It is determined that 
the strongest one-electron excitations, in decreasing order, are: 
$4d \rightarrow 4f$, $4d \rightarrow 5d$, $4f \rightarrow 5g$, $4f \rightarrow 5f$, 
$4d \rightarrow 5g$, $4f \rightarrow 6g$, $4d \rightarrow 5p$. 
The electron-impact cross-sections for $4d \rightarrow 4f$ excitation are by one 
order larger compared with  $4d \rightarrow 5d$ excitation and by two orders 
larger than $4f \rightarrow 5d$ excitation. It can be seen from  
Figs. \ref{w25_2_4}, \ref{w25_4_7}, and \ref{w25_10_30} that transitions from 
$4d^{9}4f^{4}$, $4f^{2}5p$, $4f^{2}5d$, $4f^{2}5f$, $4f^{2}5g$, and $4f^{2}6g$  
configurations feature large radiative transition probabilities. However, the 
large radiative transition probabilities alone are not enough to ensure strong 
lines in the spectrum  of \textbf{a} low-density plasma. The  distribution of 
$gA$ values will not fully reproduce the spectral shape and important 
transitions, since it does not take population mechanisms into account. 

As far as we know, no modeling of the spectra for $W^{25+}$ has been performed 
for a monoergetic electron beam, except in \cite{2007apmidf_13_45_radtke}. Their 
collisional-radiative modeling included $4f^{3}$ and $4f^{2}5l$ ($l=0,1,2,3$) 
configurations but omitted $4d^{9}4f^{4}$ and $4f^{2}5g$ configurations. 
However, the configuration interaction strengths presented in Table \ref{t2} 
show that $4f^{2}5d$ configuration strongly mixes with $4d^{9}4f^{4}$ 
configuration. Furthermore, the configuration interaction has to be taken into 
account between  $4f^{2}5f$ and $4d^{9}4f^{3}5s$ configurations. On the other 
hand, $4d^{9}4f^{4}$, $4d^{9}4f^{3}5d$, and $4f^{2}5g$ configurations have to 
be included into modeling due to their strong electron-impact excitations from 
the ground configuration.

In this work, the corona model is used to estimate population of levels and to 
find contribution of various transitions to the line formation for $W^{25+}$ in 
a low-density plasma. The population of levels from the higher levels through 
radiative cascade is  also included in our investigation. Two approaches are 
adopted to obtain electron-impact excitation rates. First of all, the 
rates are considered as being proportional to the electric multipole 
(E1, E2, E3) transition probabilities, and the corona model is used to find the 
population of levels. Then these spectra are compared with the modeling where 
the electron-impact excitation rates are calculated using the distorted-wave (
DW) approximation.

For large incident electron energies, compared with excitation energies, the 
plane-wave Born (PWB) method produces accurate electron-impact excitation rates 
but this method is restricted only to spin-allowed transitions 
\cite{Cowan_1981tass.book.....C}. On the other hand, the intermediate coupling 
mixes states with the various spins and the total spin quantum number is not 
well defined. At this energy limit, radiative transition probabilities can be 
used instead of electron-impact excitation cross-sections, because the matrix 
element of the PWB operator transforms to the matrix elements of electric 
multipole transition operators. Thus, the selection rules for the PWB 
cross-sections are identical to those for the electric multipole radiation. 
The previous study of $W^{29+}$ to $W^{37+}$  spectra in EBIT plasma 
\cite{Jonauskas2007jpb_40_2179} has demonstrated that relative line intensities 
for resonant $4p^{5}4d^{N+1} + 4p^{6}4d^{N-1}4f \rightarrow 4p^{6}4d^{N}$ 
($N=1-9$) transitions  quite well agree with the collisional-radiative modeling 
\cite{Radtke2001PhysRevA_64_012720}, but in the case when electric dipole 
transition probabilities are used in the corona model. 

In addition, the electron-impact excitation rates from the levels of the ground 
configuration are calculated within the distorted-wave approximation 
at 790 eV  of electron beam energy and $10^{12}$ cm$^{-3}$ of electron beam 
density using the FAC code. The same basis of interacting configurations is 
employed in these calculations. The Gaussian distribution function with a full 
width at half-maximum of 30 eV is used for the electron energy.


 \begin{figure}
 \includegraphics[scale=0.4]{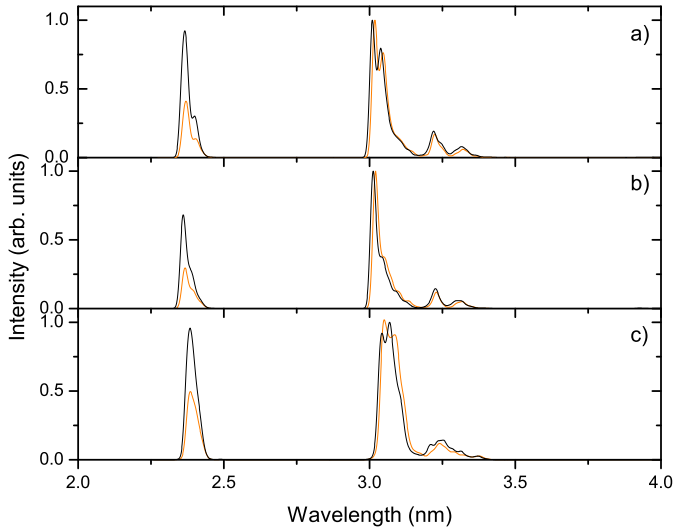}%
 \caption{\label{w25_nn_20_40} Spectra from corona models in the 2 -- 4 nm 
range when populations of levels are obtained after excitation from 
a) the ground, b) the first excited, and c) the second excited levels. 
Black line - spectrum when electric dipole transition probabilities used for 
electron-impact excitation rates; orange line - electron-impact excitation 
rates calculated using DW method. }
\end{figure}


\begin{figure}
 \includegraphics[scale=0.5]{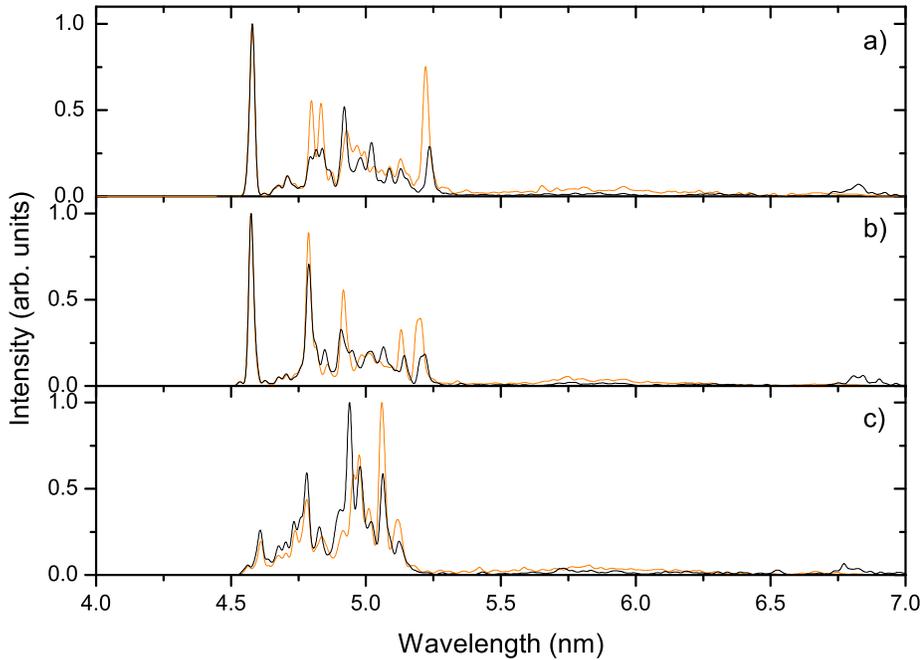}%
 \caption{\label{w25_nn_40_70} Same as Fig. \ref{w25_nn_20_40} but the 4 -- 7 
nm range. }
\end{figure}


\begin{figure}
 \includegraphics[scale=0.5]{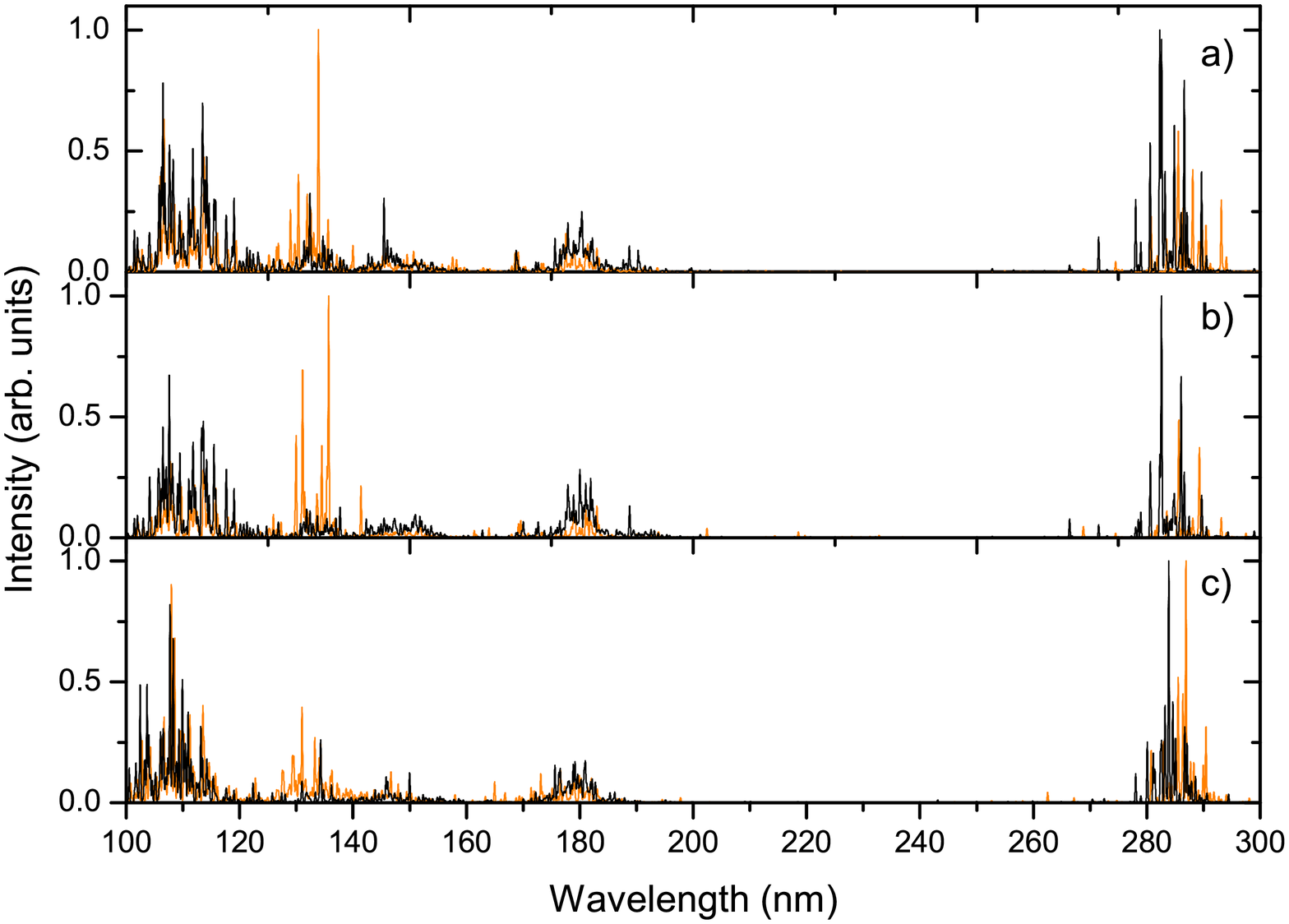}%
 \caption{\label{w25_nn_100_300} Same as Fig. \ref{w25_nn_20_40} but the 
10 -- 30 nm range. }
\end{figure}

Good agreement between both approaches can be noticed for groups of lines which 
correspond to transitions from the excited configurations $4f^{2}6g$, $4f^{2}5g$ 
(Fig. \ref{w25_nn_20_40}), $4d^{9}4f^{4}$, and $4f^{2}5d$ (Fig. 
\ref{w25_nn_40_70}) to the ground configuration. Some discrepancies between two 
applied approaches can be attributed to the different methods implemented in the 
GRASP2K and FAC codes. However, the relative intensities determined by the 
different approaches for the group of lines at 2.4 nm  differ by two times. 
It can be explained by the fact that the relative excitation rates from the 
levels of the ground configuration to the levels of $4f^{2}6g$ and $4f^{2}5g$ 
configurations differ when the PWB and DW approximations used. Spectra of $gA$ 
values show good agreement with modeled spectra too. It should be noted that the 
influence of $4f^{2}5d \rightarrow 4f^{3}$ transition to the formation of lines 
in the 4--7 nm range drastically increases in the both modeled spectra 
(Fig. \ref{w25_nn_40_70}). Many strong lines correspond to this transition while 
the corresponding $gA$ values are much smaller compared with those from the
$4d^{9}4f^{4} \rightarrow 4f^{3}$ transition.

The largest discrepancies among the spectra of two corona models and $gA$ values 
are obtained in the 10--30 nm range where $(n=5) \rightarrow (n=5)$ transitions 
are observed (Fig. \ref{w25_nn_100_300}). The radiative cascade plays a crucial 
role in the line formation for the transitions in both corona models. 
The radiative cascade has the largest influence for the lowest configurations of 
the complex.

Corona modeling, where the DW excitations are used to calculate populations of 
the excited levels, especially highlights the lines  from  the 
$4f^{2}5f \rightarrow 4f^{2}5d$ transition in the 13 nm region. This contrasts 
to the line intensities detemined using the modeling with electric transition 
probabilities. Interestingly, the intensities of the strongest lines obtained 
with the DW method are not strongly affected by radiative cascade. In the PWB 
limit, the initial configuration is mainly populated through the radiative 
cascade because direct excitations from the levels of the ground configuration 
are very weak. On the other hand, electron-impact excitation cross-sections from 
the ground configuration to the $4f^{2}5f$ configuration are quite strong 
according to our UTA calculations performed using the DW approximation. 
The same region is populated by the lines from $4f^{2}5d \rightarrow 4f^{2}5p$ 
transition. However, in this case the influence of the radiative cascade is very 
important.

The strongest increase of line intensities compared with $gA$ spectrum is 
observed in the 10--12 nm region. This region is formed by the 
$4f^{2}5s \rightarrow 4f^{3}$ transition which has not been investigated before. 
Their electric dipole transition probabilities are very small compared to other 
electric dipole transitions as it is mentioned in Sec \ref{sec:gA}. Only the 
radiative cascade from the higher levels is responsible for the strong increase 
of the population of levels of $4f^{2}5s$ configuration.

Other groups of lines at 18 nm ($4f^{2}5d \rightarrow 4f^{2}5p$, 
$4f^{2}5p \rightarrow 4f^{2}5s$) and 28 nm ($4f^{2}5p \rightarrow 4f^{2}5s$) are 
strongly affected by the radiative cascade as well. 

Finally, the most interesting result of modeling in the 10--30 nm range is a 
complex structure of lines which do not form emission bands. Those lines, close 
to 13, 18, and 29 nm, can be observed in EBIT plasma. Another interesting study 
would be in the wavelength range of the 10--12 nm  where lines from the
$4f^{2}5s \rightarrow 4f^{3}$ transitions are concentrated. These transitions 
were not previously analyzed  neither theoretically nor experimentally.

\section{Estimation of uncertainties}

Estimation of uncertainties in our data is based on the difference between 
single-configuration and CI methods. Because our CI basis includes the main 
configurations important for the considered configurations, we assume that 
the larger CI basis will not have larger impact to the energy levels, 
wavelengths, and transition probabilities. 

For $4f^{3}$ configuration, the energy levels are approximately 0.23 a.u.  
lower in CI calculations compared with the single-configuration method. Even 
better agreement is obtained for $4f^{2}5s$ (0.045 a.u.) and $4f^{2}5p$ 
(0.014 a.u.) configurations. Large mixing among the configuration state 
functions of $4d^{9}4f^{4}$ and $4f^{2}5d$ configurations indicates that these 
levels have to be investigated together. For this group of configurations, the 
disagreement between both calculations does not reach 0.2 a.u. The largest 
discrepancies appear for the more excited configurations of $n=5$ complex 
($4f^{2}5f$ - 0.312 a.u., $4f^{2}5g$ - 0.570 a.u.).

Due to different uncertainties for various configurations, the uncertainties for 
wavelengths differ for various transitions. The smallest discrepancy is 
detemined for $4f^{2}6g \rightarrow 4f^{3}$ (0.06 nm) and 
$4f^{2}5g \rightarrow 4f^{3}$ (0.12 nm) transitions. The largest discrepancies 
up to 0.5 nm take place for $4f^{2}5g \rightarrow 4f^{2}5f$ and 
$4f^{2}5f \rightarrow 4f^{2}5d$ transitions. Furthermore, the agreement between 
both methods is within 0.3 nm for $4f^{2}5d \rightarrow 4f^{2}5p$ and 
$4f^{2}5p \rightarrow 4f^{2}5s$ transitions, and within 0.2 nm for 
$4d^{9}4f^{4}+4f^{2}5d \rightarrow 4f^{3}$ transitions. 

Radiative transition probabilities obtained using single-configuration and CI 
methods do not differ by more than 30 \% for many $4f^{2}6g \rightarrow 4f^{3}$, 
$4f^{2}5g \rightarrow 4f^{3}$,  $4f^{2}5g \rightarrow 4f^{2}5f$, and 
$4f^{2}5p \rightarrow 4f^{2}5s$ transitions. Similar uncertainties can be 
attributed to transitions which include $4f^{2}5d$, and $4d^{9}4f^{4}$ 
configurations: $4f^{2}5f \rightarrow 4f^{2}5d$, $4f^{2}5d \rightarrow 4f^{2}5p$, 
and $4d^{9}4f^{4}+4f^{2}5d \rightarrow 4f^{3}$.

\section{Conclusions}

We have analyzed energy levels and radiative transition probabilities in the 
$W^{25+}$ ion using the multiconfiguration Dirac-Fock method and the extended 
basis of configurations. Our calculations show that lifetimes of some 
$4d^{9}4f^{4}$ and $4d^{8}4f^{5}$ configuration levels are comparable with the 
lifetimes of levels from the ground configuration where mostly magnetic dipole 
transitions are available. 

In addition, correlation effects have been studied using configuration 
interaction strength in the configurations responsible for the formation of the 
strongest lines in the spectrum. We have determined that the correlation effects 
are crucial for some configurations and transitions. The largest mixing of 
configurations takes place for $4f^{2}5d$, $4f^{2}5f$, and $4f^{2}5g$ 
configurations. Influence of the correlations  has been studied for the
$4f^{2}5s \rightarrow 4f^{3}$ and $4f^{2}5d \rightarrow 4f^{3}$ transitions. 
The configuration interaction allows the electric dipole transitions from the 
first excited $4f^{2}5s$ configuration to the ground $4f^{3}$ configuration. 
However, the electric dipole transition probabilities are approximately six 
orders larger than those of the electric octupole transitions.  For the 
$4f^{2}5d \rightarrow 4f^{3}$ transition, it was found that the correlation 
effects increase the electric dipole transition probabilities by an order of 
magnitude, the width of the spectrum becomes wider approximately by two times, 
and the average wavelength shifts to the shorter wavelength side. 

The corona model is used to find contribution of various transitions to the 
formation of lines in the spectrum. The modeling has demonstrated good agreement 
with the spectra of $gA$ values for strong transitions to the ground 
configuration. However, the corona modeling boosts intensities of the lines 
corresponding to the $4f^{2}5s \rightarrow 4f^{3}$ transition. Levels of the 
$4f^{2}5s$ configuration are mainly populated through radiative cascade from  
higher levels.

Two approaches are applied to estimate electron-impact excitation rates in the 
corona model -- the distorted-wave approximation and radiative transition 
probabilities used for the electron-impact excitation rates. Good agreement is 
determined between two approaches for strong transitions in the 2 -- 7 nm range. 
Modeling of spectra  in the 10--30 nm range reveals the structure of lines which 
do not merge to emission bands. We suggest further EBIT observations for those 
lines in $W^{25+}$ spectra.

\section*{Akcnowledgement}
This research was funded by European Social Fund under the Global Grant Measure 
(No.: VP1-3.1-\v{S}MM-07-K-02-015).


%

\end{document}